\documentclass[manuscript]{acmart}
\AtBeginDocument{%
  \providecommand\BibTeX{{%
    \normalfont B\kern-0.5em{\scshape i\kern-0.25em b}\kern-0.8em\TeX}}}
	
\usepackage{multirow}
\usepackage{caption}
\usepackage{subcaption}
\usepackage{makecell}
\usepackage{booktabs}
\usepackage{array}
\usepackage{bbm}
\usepackage{xcolor,colortbl}

\usepackage{ifthen}
\usepackage{mathtools}
\usepackage[export]{adjustbox}
\usepackage{xspace}
\newboolean{comments}

\definecolor{Gray}{gray}{0.9}

\newcommand{\Fig}[1]{Fig.~\ref{fig:#1}}
\newcommand{\Figure}[1]{Figure~\ref{fig:#1}}

\newcommand{\Eq}[1]{Eq.~\ref{eq:#1}}

\newcommand{\Sec}[1]{Sec.~\ref{sec:#1}}

\newcommand{\Appendix}[1]{Appendix~\ref{app:#1}}

\newcommand{\useragent}{{user agent}\xspace}
\newcommand{\interfaceagent}{{interface agent}\xspace}

\usepackage[normalem]{ulem}

\newcommand{\add}[1]{#1}
\newcommand{\addiui}[1]{#1}
\newcommand{\deliui}[1]{}

\newcommand{\del}[1]{}

\begin{document}

\begin{teaserfigure}
    \centering
    \includegraphics[width=\textwidth]{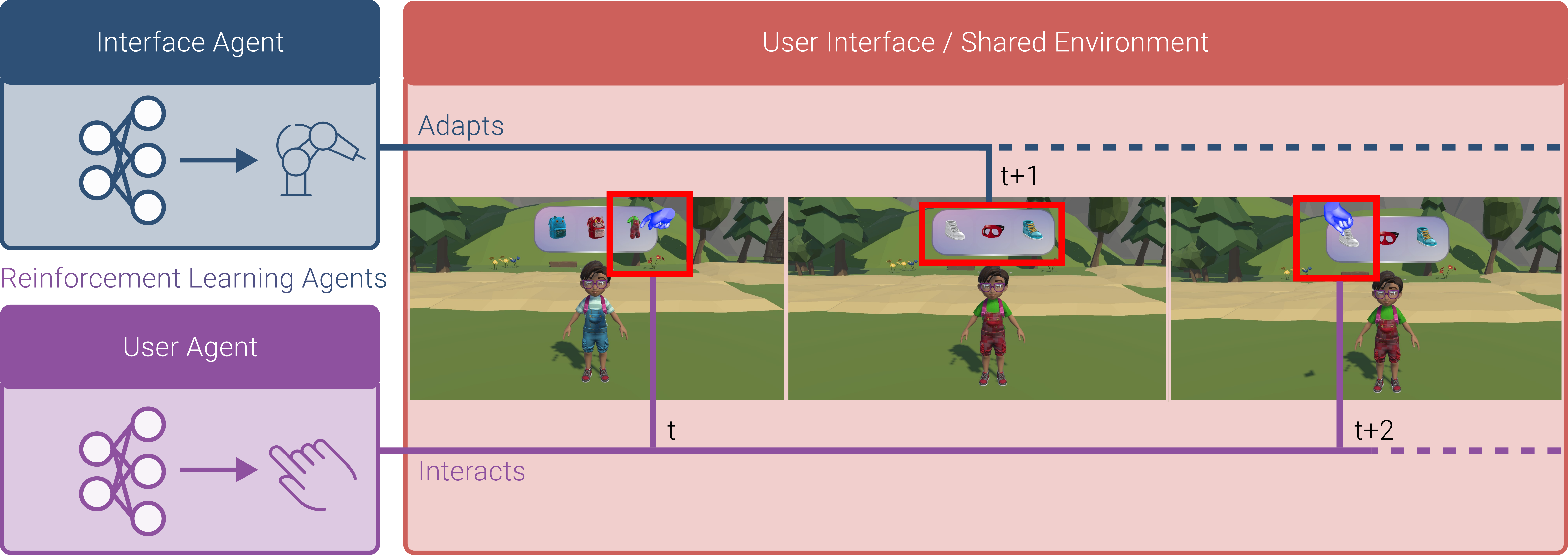}
    \caption{We formulate online user interface adaptation as a multi-agent reinforcement learning problem. Our approach comprises a user- and interface agent. The \useragent interacts with an application in order to reach a goal and the \interfaceagent learns to assist it\del{ in reaching this goal without requiring a-priori knowledge about the task}. In the depicted example the \useragent interacts with a Virtual Reality toolbar, while the \interfaceagent assigns relevant items for the \useragent. The \interfaceagent does not know the goal of the \useragent \del{but has to infer it from observing the user's actions}. Crucially, our approach does not rely on labeled offline data or application-specific handcrafted heuristics.}
    \Description{Overview figure with three components: Interface Agent, User Agent and User Interface/Shared Environment. The User-agent interacts with a menu in the environment to select clothing items. Subsequently, the interface agent adapt visible clothing items, and the user agent continues its interaction.}
    \label{fig:teaser}
\end{teaserfigure}







\title{MARLUI: Multi-Agent Reinforcement Learning for Adaptive UIs}

\newcommand{\SetOfStates}{\mathcal{S}}
\newcommand{\StatePerPolicy}{s}
\newcommand{\SetOfObservations}{\mathcal{O}}
\newcommand{\ObservationPerPolicy}{O}
\newcommand{\NumberOfPlayers}{\mathcal{N}}
\newcommand{\SetOfActions}{\mathcal{A}}
\newcommand{\ActionPerPolicy}{a}
\newcommand{\Transitions}{T}
\newcommand{\SetOfTransitions}{\mathcal{T}}
\newcommand{\ObservationTransitions}{{F}}
\newcommand{\SetOfObservationTransitions}{\mathcal{F}}

\newcommand{\SetOfRewards}{\mathcal{R}}
\newcommand{\RewardPerPolicy}{R}
\newcommand{\SetOfPolicies}{\Pi}
\newcommand{\discount}{\gamma}

\newcommand{\stack}{\mathbf{o}}
\newcommand{\error}{\mathcal{E}}
\newcommand{\action}{\mathbf{a}}
\newcommand{\observation}{\mathbf{o}}
\renewcommand{\state}{\mathbf{s}}
\newcommand{\policy}{\pi}

\newcommand{\nitems}{n_{i}}
\newcommand{\nslots}{n_{s}}

\newcommand{\pos}{\mathbf{p}}
\newcommand{\tools}{\mathbf{x}}
\newcommand{\gattr}{\mathbf{g}}
\newcommand{\menu}{\mathbf{m}}
\newcommand{\tct}{T_{task}}

\newcommand{\target}{\mathbf{t}}
\newcommand{\satweight}{\lambda}
\newcommand{\miss}{\lnot h}
\newcommand{\button}{\mathbf{b}}
\newcommand{\mt}{T_M}
\newcommand{\dect}{T_D}

\author{Thomas Langerak}
\affiliation{%
     \institution{ETH Z\"urich}
    \country{Switzerland}}
    \email{thomas.langerak@inf.ethz.ch}

\author{Sammy Christen}
\affiliation{%
     \institution{ETH Z\"urich}
    \country{Switzerland}}

\author{Mert Albaba}
\affiliation{%
     \institution{ETH Z\"urich}
    \country{Switzerland}}

\author{Christoph Gebhardt}
\affiliation{%
     \institution{ETH Z\"urich}
    \country{Switzerland}}

\author{Otmar Hilliges}
\affiliation{%
     \institution{ETH Z\"urich}
    \country{Switzerland}}

\renewcommand{\shortauthors}{Langerak, et al.}


\begin{abstract}
Adaptive user interfaces (UIs) automatically change an interface to better support users' tasks. Recently, machine learning techniques have enabled the transition to more powerful and complex adaptive UIs. However, a core challenge for adaptive user interfaces is the reliance on high-quality user data that has to be collected offline for each task. We formulate UI adaptation as a multi-agent reinforcement learning problem to overcome this challenge. In our formulation, a \useragent mimics a real user and learns to interact with a UI. Simultaneously, an \interfaceagent learns UI adaptations to maximize the \useragent's performance. The \interfaceagent learns the task structure from the \useragent's behavior and, based on that, can support the \useragent in completing its task. Our method produces adaptation policies that are learned in simulation only and, therefore, \emph{does not need real user data}. Our experiments show that learned policies generalize to real users and achieve on-par performance with data-driven supervised learning baselines.
\end{abstract}



\maketitle

\section{Introduction}
Many tasks require powerful User Interfaces (UIs). However, the more powerful the interface, the more complex it becomes. For instance, consider the plethora of factors (e.g., six degrees of freedom and real-world context) that influence the experience of Mixed Reality interfaces. One solution to the complexity problem is to dynamically adapt the UI to the user and their task by showing relevant information in a contextual and timely manner. Researchers demonstrated that adaptive UIs (AUIs) improve the usability over standard UIs in various use cases, including menus \cite{Findlater2009, Sears1994, Bailly2017, Cockburn2007}, cooperative interfaces \cite{park2018adam}, and virtual reality interfaces \cite{lindlbauer2019context}. Appropriate adaptations could help users complete their work, but designing useful adaptive UIs is challenging.

Modern AUIs \cite{Shen2009a, Shen2009b, todi2021adapting, gebhardt2019learning} predominantly leverage machine learning (ML) techniques that learn correlations between user input, user intentions, and adaptation. These approaches decrease development efforts and make AUIs more usable compared to their rule-based predecessors \cite{gebhardt2021optimal}. However, these methods rely on real user data, introducing three significant limitations. First, \addiui{existing} data is usually unavailable for the UIs of emerging technologies (data availability). Second, designers need to recollect data for any change \addiui{or iteration} in the UI design, which is an expensive and time-consuming process (design-specific data). Third, when collecting user data, it is non-trivial to ensure that the data is representative of the users' actual intentions (data quality).

We frame adapting UIs as a multi-agent reinforcement learning (MARL) problem to alleviate these challenges. In our formulation, a simulated \useragent learns to interact with a UI to complete a task. Simultaneously, we train an \interfaceagent that learns to adapt the same UI to help the \useragent achieve the task more efficiently. By working in a simulated environment, our approach does not rely on \addiui{tediously} collected user data (data availability) since the \useragent learns to use the interface through online interactions and thus generates unambiguous data (data quality). Finally, our method implements a general formulation of the interface adaptation problem (i.e., the method is general agnostic) and can learn meaningful policies for different interfaces and tasks (design-specific data).

Specifically, we model the \useragent as a Hierarchical Reinforcement Learning (HRL) agent that learns to navigate an interface and complete its task by interacting with the UI. To achieve realistic behavior, we decompose high-level decision-making (e.g., the decision to select a menu item) from motor control (e.g., moving the cursor to the corresponding menu slot). The \interfaceagent is a single reinforcement learning (RL) policy. Its goal is to assist the \useragent in completing the task more efficiently, for instance, by assigning the correct items to a toolbar with limited slots in a game character creation task (cf. \Fig{teaser}). Crucially, the \interfaceagent does not require access to the \useragent's intent. Instead, the \interfaceagent learns the underlying task structure. In our setting, both the \useragent and the \interfaceagent share the same reward for completing the task efficiently and accurately.

To demonstrate the feasibility of our method, we introduce four proof-of-concept use cases in VR: a game character design tool with an intelligent toolbar, an intelligent numeric keypad, a block tower building game, and an application that helps to select virtual objects that are out of reach.

Our main goal is to show that our method is competitive with baselines that require carefully collected real user data. Not relying on data would make our method suitable for a large variety of interfaces and tasks. To evaluate our goal, we perform both an evaluation with real users and in silico studies in the character creation task. Our study with real users compares our method against different data-driven baselines. We find that training the \interfaceagent with our simulated \useragent transfers well to humans and performs on par with previous data-driven methods regarding task completion time. 

In summary, this paper makes four key contributions: (1) a MARL-based framework to adapt user interfaces online without relying on real user data; (2) a Hierarchical Reinforcement Learning-based, cognitively inspired \useragent that can learn to operate a user interface and enables an \interfaceagent to learn adaptations that are useful to real end-users; (3) a goal-agnostic \interfaceagent that learns the underlying structure of the task purely by observing the action in the interface by the \useragent; and (4) empirical results showing the effectiveness of our approach and four different usage scenarios to demonstrate its general applicability.


\section{Related Work}
\label{sec:related}
This paper proposes multi-agent RL as a framework for adaptive UIs. Our method features a \useragent that models human interaction behavior and an \interfaceagent that adapts the UI to support the \useragent.
Most related to our work is research on computational user modeling, methods for adaptive UIs, and (MA)RL.

\subsection{Computational User Modelling}
Computational user modeling has a long tradition in HCI. These models predict user performance and are essential for UI optimization \cite{oulasvirta2018computational}. Early work relies on heuristics \cite{card1986model, card1980klm, card1983the, kieras1997overview, anderson1997act} and on simple mathematical models \cite{fitts1954information, hick1952rate}.
More recent work extends these models and, for instance, predicts the operating time for a linear menu \cite{10.1145/1240624.1240723}, gaze patterns \cite{salvucci2001integrated}, or cognitive load \cite{duchowski2018index}.

Recently, reinforcement learning gained popularity within the research area of computational user models.
This popularity is due to its neurological plausibility \cite{botvinick2012hierarchical, frank2012mechanisms}, allowing it to serve as a model of human cognitive functioning.
The underlying assumption of RL in HCI is that users behave rationally within their bounded resources \cite{gershman2015computational, oulasvirta2022computational}.
There is evidence that humans use such strategy across domains, such as in causal reasoning \cite{denison2013rational} or perception \cite{gershman2012multistability}.
In human-computer interaction, researchers have leveraged RL to automate the sequence of user actions in a KLM framework \cite{leino2019computer} or to predict fatigue in volumetric movements \cite{cheema2020predicting}.
It was also used to explain search behavior in user interfaces \cite{yang2020predicting} or menus \cite{chen2015emergence} and as a model for multitasking \cite{jokinen2021multitasking}.
Most similar to our work is research on hierarchical reinforcement learning for user modeling.
\citeauthor{jokinen2021touchscreen} \cite{jokinen2021touchscreen} show that human-like typing can emerge with the help of Fitts' Law and a gaze model.
Other works show that HRL can elicit human-like behavior in task interleaving \cite{gebhardt2020hierarchical} or touch interactions \cite{jokinen2021touchscreen}.
Inspired by this work, we design our user agent by decomposing high-level decision-making from motor control.
Using two hierarchical levels yields simulated behavior.

\subsection{Methods for Adaptive UIs}
UI adaptation can either be offline, to computationally design an interface, or online, to adapt the UI according to users' goals. We will focus on online adaptive UIs and refer readers to \cite{combinatorialoptimizationdesign2020oulasvirta, functionalityselection2017oulasvirta} for an overview of computational UI design.

\subsubsection{Heuristics, Bayesian Networks \& Combinatorial Optimization}
In early works, heuristic- or knowledge-based approaches are used to adapt the UI  \cite{Browne1990,Stephanidis1997,Smith2010}.
Similarly, multi-agent systems employ rule-based and message-passing approaches  \citep{Rich1998,Rich2005,Yorke2012}. Another popular technique for AUIs is domain-expert-designed Bayesian networks  \citep{Horvitz1998, Bosma2004}.
More recently, combinatorial optimization was used to adapt interfaces dynamically \cite{park2018adam, lindlbauer2019context}.
The downside of these approaches is that experts need to specify user goals using complex rule-based systems or mathematical formulations. 
Creating them comprehensively and accurately requires developers to foresee all possible user states, which is tedious and requires expert knowledge.
Commonly, these approaches also get into conflict when multiple rules or objectives apply.
This conflict often results in unintuitive adaptations.
In contrast, our method only requires the layout of the UI.
From its representation as an RL environment, we learn policies that meaningfully adapt the UI and realistically reproduce user behavior.

\subsubsection{\del{Machine Learning} \add{Supervised Learning}}
Leveraging machine learning can overcome the limitations of heuristic-, network-, and optimization-based systems by learning appropriate UI adaptions from user data.
Traditional machine learning approaches commonly learn a mapping from user input to UI adaptation. 
Algorithms like nearest neighbor  \cite{Maes1995, Lashkari1997}, Na\"ive Bayes  \cite{McCreath2006,Faulring2010}, perceptron \cite{Shen2009a, Shen2009b}, support vector machines \cite{Berry2011}, or random forests \cite{Pejovic2014, Mehrotra2015} are used and models are learned offline \cite{Berry2011} and online \cite{Shen2009a}.
Due to the problem setting, these approaches require users' input to be highly predictive of the most appropriate adaptation. 
Furthermore, it restricts the methods to work in use cases where myopic planning is sufficient, i.e., a single UI adaptation leads users to their goal. 
In contrast, our method considers multiple goals when selecting an adaptation and can lead users to their goal using sequences of adaptations. 

More recent work overcomes the limitations stemming from simple input-to-adaptation mapping by following a two-step approach. They (1) infer users' intention based on observations and (2) choose an appropriate adaptation based on the inferred intent \cite{oulasvirta2018computational}.
Such work uses neural networks, and user intention is modeled either explicitly \cite{kolekar2010learning, soh2017deep} or as a low-dimensional latent representation \cite{RIZZOGLIO2021}.
However, these approaches are still highly dependent on \deliui{the quality of the} training data, which may not even be available for emerging technologies. \deliui{In contrast, our method does not depend on pre-collected user data \emph{and} can learn supportive policies just by observing simulated user behavior. } \addiui{In contrast, our method can learn supportive policies without pre-collected user data by just observing simulated user behavior.}

\subsubsection{Bandits \& Bayesian Optimization} 
Bandit systems are a probabilistic approach often used in recommender systems \cite{glowacka2019bandit}.
In a multi-armed bandit setting, each adaptation is modeled as an arm with a probability distribution describing the expected reward.
The Bayes theorem updates the expectation, given a new observation and prior data.
Related work leverages this approach for AUIs \cite{lomas2016interface,koch2019may,kangas2022scalable}.
Bayesian optimization is a sample-efficient global optimization method that finds optimal solutions in multi-dimensional spaces by probing a black box function \cite{shahriari2015taking}. In the case of AUIs, it is used to find optimal UI adaptations by sampling users' preferences \cite{Koyama2014, Koyama2016}.
Both approaches trade off exploration and exploitation when searching for appropriate adaptations (i.e., exploration finds entirely new solutions, and exploitation improves existing solutions), rendering them suitable approaches to the AUI problem.

However, such methods are not able to plan adaptations over a sequence of interaction steps, i.e., they plan myopic strategies.
In addition, these approaches need to sample user feedback to learn or optimize for meaningful adaptations and, hence, also rely on high-fidelity user data.
Furthermore, as users themselves learn during training or optimization, solutions can converge to sub-optimal user behavior as such methods reduce exploration with convergence. 
In contrast, our method can plan adaptations over a sequence of interaction steps learned from realistic, simulated user data.  

\subsubsection{Reinforcement Learning} 
Reinforcement learning is a natural approach to solving the AUI problem, as its underlying decision-making formalism implicitly
captures the closed-loop iterative nature of HCI \cite{howes2018interaction}. 
It is a generalization of bandits and learns policies for longer horizons, where current actions can influence future states.
This generalization enables selecting UI adaptations according to user goals that require multiple interaction steps.
Its capability makes RL a powerful approach for AUIs with applications in dialog systems \cite{Gasic2014, Su2017}, crowdsourcing \cite{PengCrowdsourcing2013, Hu2018}, sequential recommendations \cite{Chen2019, Liu2018, Liebman2015}, information filtering \cite{seo2000reinforcement}, personalized web page design \cite{ferretti2014exploiting}, and mixed reality \cite{gebhardt2019learning}.
Similar to our work is a model-based RL method that optimizes menu adaptations \cite{todi2021adapting}.

Current RL methods sample predictive models \cite{todi2021adapting, Gasic2014, Hu2018} or logged user traces  \cite{gebhardt2019learning}. However, these predictive models and offline traces represent user interactions with non-adaptive interfaces. Introducing an adaptive interface will change user behavior; so-called co-adaptation \cite{mackay2000responding}. Hence, it is unclear if the learned model can choose meaningful adaptations when user behavior changes significantly due to the model's introduction. In contrast, our user agent learns to interact with the adapted UI; hence, our interface agent learns on behavioral traces from the adapted setting.

\subsubsection{Multi-Agent Reinforcement Learning}
MARL is a generalization of RL in which multiple agents act, competitively or cooperatively, in a shared environment \cite{zhang2021multi}.  
Multi-agent systems are common in games \cite{baker2019emergent, jaderberg2019quake}, robotics \cite{ota2006multiagent,sariff2018multiagent}, or modeling of social dilemmas \cite{chao2015social,leibo2017multi}. MARL is challenging since multiple agents change their behavior as training progresses, making the learning problem non-stationary. Common techniques to address this issue is via implicit \cite{tian2020implicit} or explicit \cite{foerster2016learning} communication, centralized critic functions \cite{lowe2017multi,yu2021surprising}, or curricula \cite{epciclr2020,wang2020curriculum}. We take inspiration from the latter and use a curriculum during the training of our agents.

Closest to our work is \cite{debard2020multiagent}, which proposes a multi-agent system that maps 2D interface trajectories to actions for navigating 3D virtual environments. A user agent learns interactions on a 2D interface. A decoder that is trained on real user data maps the user agent's actions to 2D touch gestures. A second agent then translates these 2D touch gestures into 3D operations. 
In their setting, the interface agent does not observe the environment itself but receives the actions of the user agent as its state.
Our work extends their setting to the case where the user agent and the interface agent observe and manipulate the same UI. Furthermore, we do not rely on real world user data.

\section{Multi-Agent Reinforcement Learning for Adaptive UIs}
Adaptive UIs aim to automatically and intelligently adapt a UI to guide the user toward completing their task. Thus, they usually require logged user data or online user feedback.
This section explains how our proposed method, MARLUI, can create AUIs without needing data and provides an intuition on its learning procedure.

\begin{figure*}[t]
    \centering
    \includegraphics[width=0.9\textwidth]{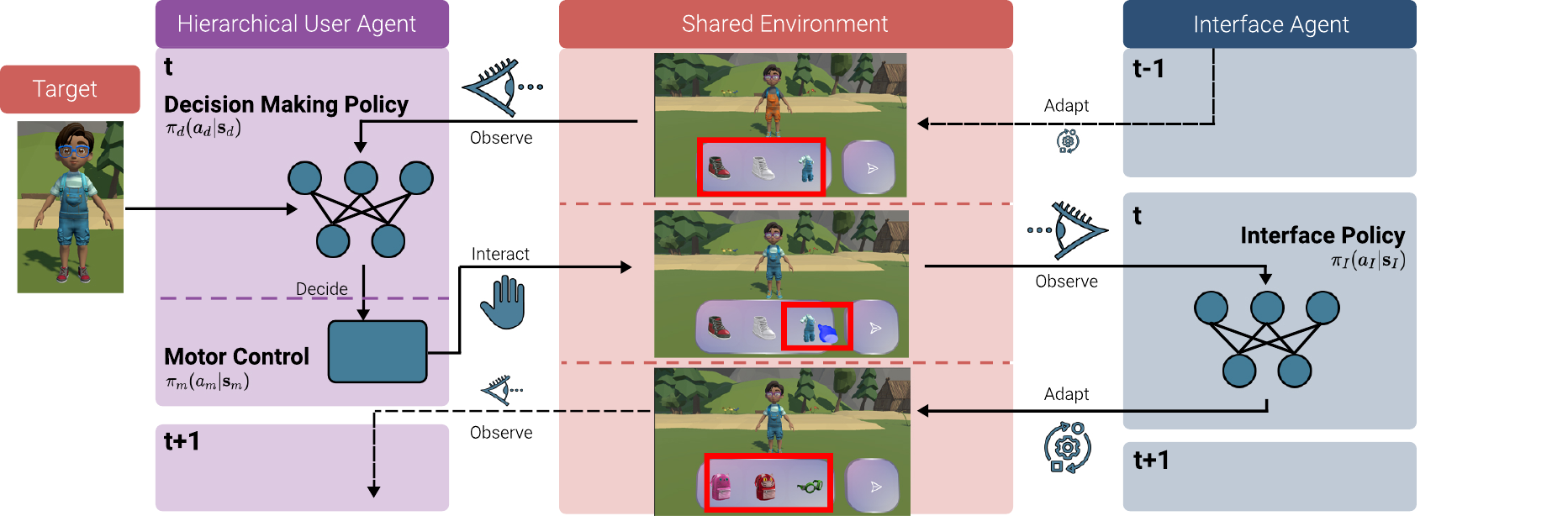}
    \caption{Our \interfaceagent and \useragent act in the same environment. The \useragent is modeled as a two-level hierarchy with a high-level decision-making policy $\policy_d$ and a low-level motor control policy $\policy_m$. The agent interacts with the UI. The high-level agent observes that state of the environment (\Eq{sd}) and chooses a specific menu slot as target accordingly (\Eq{ad}). The lower level receives this action and computes a movement (\Sec{ll}). The \interfaceagent policy $\policy_I$ adapts the interface to assist the \useragent in achieving its task more efficiently. It observes user actions in the UI (\Eq{si}) and decides on adaptations. Note that the \interfaceagent cannot access the goal, making the problem partially observable.}
    
    \Description{
        Flow Chart with three columns labeled Hierarchical User Agent, shared Environment, and Interface Agent.
        The shared environment contains a GUI with different selection possibilities for different clothing. The hierarchical user agent is subdivided into 1) decision-making policy. This observes the GUI and outputs an action. 2) A motor control policy, this takes as input the action and interacts subsequently with the GUI. The interface agent has a single component. This component observes the user interaction and adapts the GUI.  
    }
    \label{fig:overview}
\end{figure*}

The MARLUI framework consists of three main components: a \useragent, an \interfaceagent, and a user interface as a shared environment (see \Fig{overview}).
A simulated \useragent learns to achieve a goal in a simulated UI environment while the \interfaceagent, which is the AUI component, learns to adapt the interface to support the user better.
We build on the idea that by providing a model of user and UI that are realistic enough given its real-world counterparts, the \interfaceagent can provide meaningful support to real users in the real task.
 This becomes clearer if we consider MARLUI's learning procedure: Before starting with training, we specify the goal the \useragent should achieve by interacting with the UI. The goal is randomly sampled and unknown to the \interfaceagent. During training, the \useragent tries to reach this state through trial and error. If the \useragent behaves similarly enough to a human, then the \interfaceagent gets meaningful observations to learn from. 
In the case of the toolbar in \Figure{teaser}, the \useragent clicks through random sequences of menu items and then observes if the result matches the goal state. The \useragent will explore paths in the UI and adapt its behavioral policy based on the degree to which the path leads to success. To learn \useragent policies that exhibit human-like behavior, we consider task-relevant factors of human functioning in its model (e.g., human motor control).

At the same time and in the same environment (i.e., the same UI), we train an \interfaceagent. While the goal of the \useragent is to reach a goal state, the \interfaceagent aims to change the UI such that the \useragent can achieve its goal more efficiently. In the toolbar example, it would randomly assign items to menu slots and observe if this helps the \useragent to make its desired changes on the game character.
In an exploration-exploitation manner, the \interfaceagent and \useragent jointly explore the state space of the problem. 
From past observations of the \useragent interactions with the UI, the \interfaceagent will learn which goal the \useragent attempts to achieve and how to guide it towards completing it.
This avoids the need for the \interfaceagent to have direct access to the goal state, i.e., it makes it goal-agnostic. 

In the following, we formally introduce reinforcement learning and describe the details of our method. We then show the extent to which our approach generalizes to real users.

\section{Background}
We briefly introduce (multi-agent) reinforcement learning and its underlying decision processes. Specifically, we assume that users behave according to Computational Rationality (CR) \cite{oulasvirta2022computational}. This allows us to frame user behavior as a Partially Observable Markov Decision Process (POMDP).

\subsection{Partially Observable Markov Decision Processes}
Partially Observable Markov Decision Processes (POMDP) is a mathematical framework for single-agent decision-making in stochastic partially observable environments \cite{aastrom1965optimal}, which is a generalization over Markov Decision Processes \cite{howard1960dynamic}. A POMDP is a seven-tuple ($\StatePerPolicy, \ObservationPerPolicy, \ActionPerPolicy, \Transitions, \RewardPerPolicy, \gamma$), where $\StatePerPolicy$ is a set of states, $\ObservationPerPolicy$ is set of observations and $\ActionPerPolicy$ a set of actions. In POMDPs, the exact states ($s \in \StatePerPolicy$) of the evolving environment may or may not be captured fully. Therefore, observations ($o \in \ObservationPerPolicy$) represent the observable states, which may differ from the exact state. $\Transitions: \StatePerPolicy \times \ActionPerPolicy \times \StatePerPolicy \rightarrow [0,1]$ is a transition probability function, where $\Transitions (\state', \action, \state)$ is the probability of the transition from state $\state'$ to $\state$ after taking action $\action$. Similarly, $\ObservationTransitions: \StatePerPolicy \times \ActionPerPolicy \times \ObservationPerPolicy \rightarrow [0,1]$ is an observation probability function, where $\ObservationTransitions (\observation, \action, \state')$ is the probability of observing $\observation$ while transitioning to $\state'$ after taking action $\action$. $\RewardPerPolicy : \StatePerPolicy \times \ActionPerPolicy \rightarrow \mathbb{R}$ is the reward function, discounted with factor $\gamma$.

\subsection{Reinforcement Learning}
Reinforcement Learning is a machine learning paradigm that rewards desired and penalizes undesired behavior. Generally, an agent observes an environment state and tries to take optimal actions to maximize a numerical reward signal. A key difference with supervised learning is that RL learns through exploration and exploitation rather than from an annotated dataset. 

We follow the standard formulation of RL as an MDP \cite{sutton1998introduction}, but use observations rather than states since the problem we are working on is partially observable. In our setting the observation space is a subspace of the state space (the \interfaceagent does not have access to the internal state of the \useragent), and observations are deterministic: $\ObservationTransitions(\observation,\action,\state') = 1$. The goal is to find an optimal policy $\policy:\ObservationPerPolicy \rightarrow \ActionPerPolicy$, a mapping from observations to actions that maximizes the expected return: $\mathbb{E}\left[ \sum_{t=0}^T \discount^t\RewardPerPolicy_i(\observation_t,\action_t)\right ]$. 
Since both observation- and action spaces can be high-dimensional, neural networks are used for policy learning (i.e., we approximate the policy as $\policy_{\theta}$, where $\theta$ are the learned parameters). 

\subsection{Multi-Agent Reinforcement Learning}
Standard reinforcement learning formulations built upon MDPs or POMDPs assume a single policy. Stochastic games generalize MDPs for multiple policies \cite{shapley1953stochastic}. When players do not have perfect information about the environment, stochastic games become partially observable stochastic games. A partially observable stochastic game is defined as an eight-tuple $\left(N, \SetOfStates, \SetOfObservations, \SetOfActions, T, \SetOfObservationTransitions, \SetOfRewards, \gamma \right )$, where $N$ is the number of policies. $\SetOfStates = \StatePerPolicy_1 \times ... \times \StatePerPolicy_N$ is a finite set of state sets, and $\SetOfObservations = \ObservationPerPolicy_1 \times ... \times \ObservationPerPolicy_N$ is a finite set of observation sets, with subscripts indicating different policies. $\SetOfActions = \ActionPerPolicy_1 \times ... \times \ActionPerPolicy_N$ is a finite set of action sets. $T$ is the transition probability function. $\SetOfObservationTransitions = \ObservationTransitions_1 \times ... \times \ObservationTransitions_N$ defines a set of observation probability functions of different players. A set of reward functions is defined as $\SetOfRewards = \RewardPerPolicy_1, ... \RewardPerPolicy_N $. Furthermore, we define a set of policies as $\SetOfPolicies = {\policy_1, ... \policy_N}$. Finally, $\gamma$ is the discount factor.  

All policies have their individual actions, states, observations, and rewards. In this paper, we optimize each policy individually, while the observations are influenced by each other's actions. We use model-free RL (for comparison to model-based RL, see \cite{polydoros2017survey}). This set of algorithms is used in an environment where the underlying dynamics $\Transitions (\state', \action, \state)$ and $\ObservationTransitions (\observation, \action, \state')$ are unknown, but it can be explored via trial-and-error. In the method section, we use the terms state and observation interchangeably.

\section{Method}
We present a general task description and outline the model of our \useragent and \interfaceagent (\Fig{overview}).

\subsection{General Task Description}
We model tasks to be completed if the user achieves their desired goal. For game character creation, a goal can be the desired configuration of a character with a certain shirt (red, green, blue) or backpack (pink, red, blue). We represent the goal as a one-hot vector encoding $\gattr$ of these attributes. A one-hot vector can be denoted as  $\mathbb{Z}_{2}^{j}$, where $j$ is the number of items. For the previous example, $\gattr$ would be in $\mathbb{Z}_2^6$ as it possesses six distinct items. 

Furthermore, the \useragent can access an input observation denoted by $\tools$. For example, this can correspond to the current character configuration. The current input observation, $\tools$, and the goal state $\gattr$ are identical in dimension and type. 

The \useragent interacts with the interface and attempts to make the input observation and goal state identical as fast as possible, such that $\tools = \gattr$. Each interaction updates $\tools$ accordingly, and a trial terminates once they are identical. In the character creation example, this would be the case if the shirt and backpack of the edited character are identical to the desired configuration. The \interfaceagent makes online adaptations to the interface. It does \emph{not} know the specific goal of a user. Instead, it needs to observe user interactions with the interface to learn the underlying task structure that will yield the optimal adaptations, e.g., the user likely wants to configure the backpack after configuring the shirt.

\subsection{User Agent}
\label{sec:user_agent}
First, we introduce the \useragent, which interacts with an environment to achieve a certain goal (e.g., select the intended attributes of a character). The agent tries to accomplish this as fast and accurately as possible. Thus, the \useragent first has to compare the goal state and input observation and then plan movements to reach the target. We model the user as a hierarchical agent with separate policies. 
Specifically, we introduce a two-level hierarchy: a high-level decision-making policy $\policy_d$ that computes a target for the agent (high-level decision-making), and a Fitts'-Law-based low-level motor policy $\policy_m$ that decides on a strategy to reach this target. \add{We approximate visual cost with the help of existing literature}. We now explain both policies in more detail.

\subsubsection{High-level Decision-Making Policy}
The high-level decision-making policy of the hierarchy is responsible \add{to select the next target item in the interface}\del{for high-level decisions, i.e., what is the next target}. The overall goal of the policy is to complete a given task while being as fast as possible. Its actions are based on the current observation of the interface, the goal state, and the agent's current state. More specifically, the high-level state space $\StatePerPolicy_d$ is defined as:

\begin{equation}
    \StatePerPolicy_d = \left (\pos, \menu, \tools, \gattr \right ),
    \label{eq:sd}
\end{equation}
which comprises: i) the current position of the \useragent's end-effector \add{normalized by the size of the UI}, $\pos \in I^n$ (where $n$ denotes the dimensions, e.g., 2D vs 3D), ii) an encoding of the assignment of each item $\menu \in \mathbb{Z}_2^{\nitems \times \nslots}$, with $\nitems$ and $\nslots$ being the number of menu items and environment locations, respectively, iii) the current input state $\tools \in \mathbb{Z}_2^{\nitems}$, and iv) the goal state $\gattr \in \mathbb{Z}_2^{\nitems}$. Here, $I$ denotes the unit interval $[0,1]$, and 
$\mathbb{Z}_{2}^{n}$ is the previously described set of integers.
The item-location encoding $m$ represents the current state of a UI. It can be used, for instance, to model item-to-slot assignments. The action space $\ActionPerPolicy_D$ is defined as:
\begin{equation}
    \ActionPerPolicy_d = \target,
    \label{eq:ad}
\end{equation}
which indicates the next target slot $\target \in \mathbb{N}_{\nslots}$. The reward for the high-level decision-making policy consists of two weighted terms to trade-off between task completion accuracy and task completion time: i) how different the current input observation $\tools$ is from the goal state $\gattr$, and ii) the time it takes to execute an action. Therefore, the high-level policy needs to learn how items correlate with the task goal as well as how to interact with any given interface. With this, we define the reward as follows: 

\begin{equation}
    \RewardPerPolicy_d =  \alpha \underbrace{\error_{gd}}_{i)} - (1-\alpha)\underbrace{\left(\dect + \mt\right)}_{ii)} + \mathbbm{1}_{\text{success}},
    \label{eq:rd}
\end{equation}
where $\error_{gd}$ is the difference between the input observation and the goal state, $\alpha$ a weight term, $\mt$  the movement time as an output of the low-level policy, $\dect$ the decision time, and $\mathbbm{1}_{\text{success}}$ an indicator function that is 1 if the task has been successfully completed and 0 otherwise. 

In addition to movement time, we also need to determine the decision time $\dect$. To this end, we are \addiui{inspired by} the SDP model \cite{10.1145/1240624.1240723}. This model interpolates between an \addiui{approximated} linear visual search-time component ($T_s$) and the Hick-Hyman decision time \cite{hick1952rate} ($T_{hh}$), \addiui{both are functions that take into account the number of menu items and user parameters}. We refer to \cite{10.1145/1240624.1240723} for more details. 

We define the difference  $\error_{gd}$ between the input observation $\tools$ and the goal state $\gattr$ as the number of mismatched attributes:
\begin{equation}
    \error_{gd} = - \sum_{x \in \gattr, y \in \tools }\frac{\mathbbm{1}_{x \neq y}}{n_{attr}},
\end{equation}
where $\mathbbm{1}$ is an indicator function that is $1$ if $x\neq y$ and else $0$,  $x$ and $y$ are individual entries in the vectors $\gattr$ and $\tools$ respectively, and $n_{attr}$ is the number of attributes (e.g., shirt, backpack, and glasses).

\subsubsection{Low-Level Motor Control Policy}
\label{sec:ll}
The low-level motor control policy is a non-learned controller for the end-effector movement. In particular, given a target, it selects the parameters of an endpoint distribution (mean $\mu_\pos$ and standard deviation $\sigma_{\pos}$) . We set $\mu_\pos$ to the center of the target. The target $\target$ is the action of the higher-level decision-making policy ($\ActionPerPolicy_D$). 
\add{Following empirical results \cite{fitts1954information},} we set $\sigma_{\pos}$ to 1/6th of a menu slot width to reach a hitrate of 96\%.

Given the current position and the endpoint parameters (mean and standard deviation), we compute the predicted movement time using the WHo Model \cite{guiard2015mathematical}.
\begin{equation}
        \mt = \left ( \frac{k}{(\sigma_{\pos}/d_{\pos}-y_0)^{1-\beta}} \right ) ^{1/\beta} + \mt^{(0)},
\end{equation}
where $k$ and $\beta$ are parameters that describe a group of users, $\mt^{(0)}$ is the minimal movement time, and $y_0$ is equal to the minimum standard deviation. The term $d_{\pos}$ indicates the traveled distance from the current position to the new target position $\mu_\pos$. We follow literature for the \add{values of} other parameters \cite{guiard2015mathematical, jokinen2021touchscreen}. We sample a new position from a normal distribution: $\pos \sim \mathcal{N}\left(\mu_{\pos}, \sigma_{\pos}\right)$.

\subsection{Interface Agent}
The \interfaceagent makes \add{discrete} changes to the UI to maximize the performance of the \useragent. 
For instance, it assigns items to a toolbar to simplify their selection for the \useragent.
Unlike the \useragent, we model the \interfaceagent as a flat RL policy. The state space $\StatePerPolicy_I$ of the interface agent is defined as:

\begin{equation}
    \StatePerPolicy_I = \left (\pos, \tools, \menu, \stack \right ),
    \label{eq:si}
\end{equation}

which includes: i) the position of the user $\pos \in I^2$, ii) the \add{input observation}\del{current state of the tool} $\tools \in \mathbb{Z}_2^{\nitems}$, iii) the current state of the \add{UI}\del{menu} $\menu \in \mathbb{Z}_2^{\nitems \times \nslots}$, and iv) a vector including the history of interface elements the \useragent interacted with (commonly referred to as stacking). The action space $\ActionPerPolicy_I \in \mathbb{Z}$ and its dimensionality is application-specific. 
The goal of the \interfaceagent is to support the \useragent. Therefore, the reward of the \interfaceagent is directly coupled to the performance of the \useragent. We define the reward of the \interfaceagent to be equal to the reward of the \useragent's high-level policy:

\begin{equation}
    \RewardPerPolicy_{I} = \RewardPerPolicy_{D}.
    \label{eqn:ri}
\end{equation}

\add{
Note that the \interfaceagent does \emph{not} have access to the \useragent's goal $\gattr$ or target $\target$.
To accomplish its task, the \interfaceagent needs to learn to help the \useragent based on an implicit understanding of i) the objective of the \useragent, and ii) the underlying task structure. Our setting allows the \interfaceagent to gain this understanding solely by observing the changes in the interface as the result of the \useragent's actions. This makes the problem more challenging but also more realistic.}

\section{Implementation}
We train the user and interface agents' policies simultaneously. All policies receive an independent reward, and the actions of the policies influence a shared environment. We execute actions in the following order: (1) the interface agent's action, (2) the user agent's high-level action, followed by (3) the user agent's low-level motor action. The reward for the two learned policies is computed after the low-level motor action has been executed. The episode is terminated when the \useragent has either completed the task or exceeded a time limit.

We implement our method in Python 3.8 using RLLIB \cite{liang2018rllib} and Gym \cite{brockman2016openai}. We use PPO \cite{schulman2017proximal} to train our policies. We use 3 cores on an Intel(R) Xeon(R) CPU @ 2.60GHz during training. Training takes $\sim$36 hours. We utilize an NVIDIA TITAN Xp GPU for training. The user agent's high-level decision-making policy $\policy_d$ is a 3-layer MLP with 512 neurons per layer and ReLU activation functions. The  interface agent's policy  $\policy_I$ is a two-layer network with 256 neurons per layer and ReLU activation functions. \addiui{We sample the full state initialization (including goal) from a uniform distribution. We use stochastic sampling for our exploration-exploitation trade-off.} We use curriculum learning to increase the task difficulty and improve learnability; for more information, see \Appendix{curriculum}. The difference between agents of different applications is their respective state- and action spaces and the set of goals.
\section{Evaluation}
\begin{figure}
    \centering
    \includegraphics[width=0.6\columnwidth]{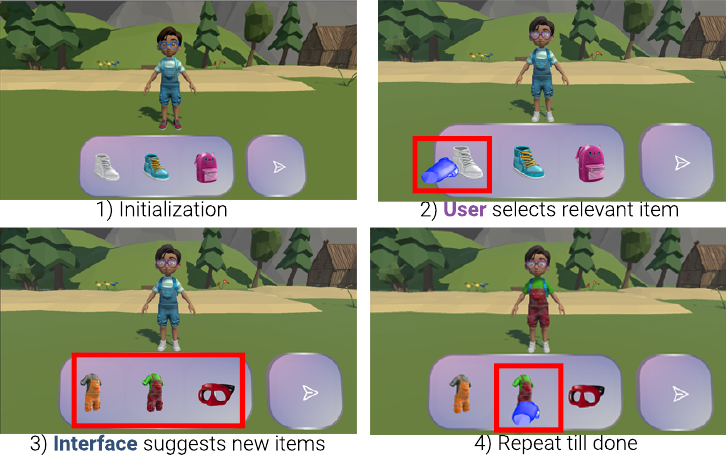}
    \caption{In our proposed task, the user (agent) matches a game character selection to a target state  (1). The user operates a toolbar with three slots (2). The interface agent assigns the most relevant items to the available slots (3). This cycle continues till the two characters match (4).}
    \Description{Task explanation figure with four subfigures. The first figure shows the initalization of the UI. In the second figure, user selects a clothing item and in the third figure interface adapts UI to suggest new items. In the fourth figure, it is showed that this process continues until the task is completed.}
    \label{fig:task}
\end{figure}

MARLUI aims to learn UI adaptations from simulated users that can support real users in the same task.
-Specifically, we want our method to produce AUIs that are competitive with baselines that require carefully collected real user data.
In this section, we evaluate if our approach achieves this goal. Thus, we first conduct an \emph{in silico study} to analyze how the \interfaceagent and the \useragent solve the UI adaptation problem in simulation. Then, we perform a \emph{user study} to investigate if policies of the \interfaceagent that were learned in simulation generalize to real users.

\subsection{Task \& Environment}
\label{sec:task}

To conduct the evaluation, we introduce the \emph{character-creation task} (see \Fig{task}).
In this task, a user creates a virtual reality game character by changing its attributes. A character has five distinct attributes with three items per attribute: i) shoes (red, blue, white), ii) shirt (orange, red, blue), iii) glasses (reading, goggles, diving), iv) backpack (pink, blue, red), and v) dance (hip hop, break, silly). The characters' attribute states are limited to one per attribute, i.e., the character cannot be dancing hip hop and break simultaneously. This leads to a total of 15 attribute items and 243 character configurations. We sample uniformly from the different configurations. 

The game character's attributes can be changed by selecting the corresponding items in a toolbar-like menu with three slots. The user can cycle through the items by selecting "Next."
The static version of the interface has all items of an attribute assigned to the three slots, and every attribute has its own page (e.g., all shoes, if the user presses next, all backpacks). 
The character's attribute states correspond to the current input state $\tools$ and the target state $\gattr$, where $\gattr$ is only known to the \useragent. The goal of the \interfaceagent to reduce the number of clicks necessary to change an attribute, by assigning the relevant items to the available menu slots. For the \emph{user agent}, the higher level selects a target slot, and the lower level moves to the corresponding location. 

\subsection{In Silico}
\paragraph{Training}

\begin{figure}
    \centering
    \includegraphics[width=0.6\columnwidth]{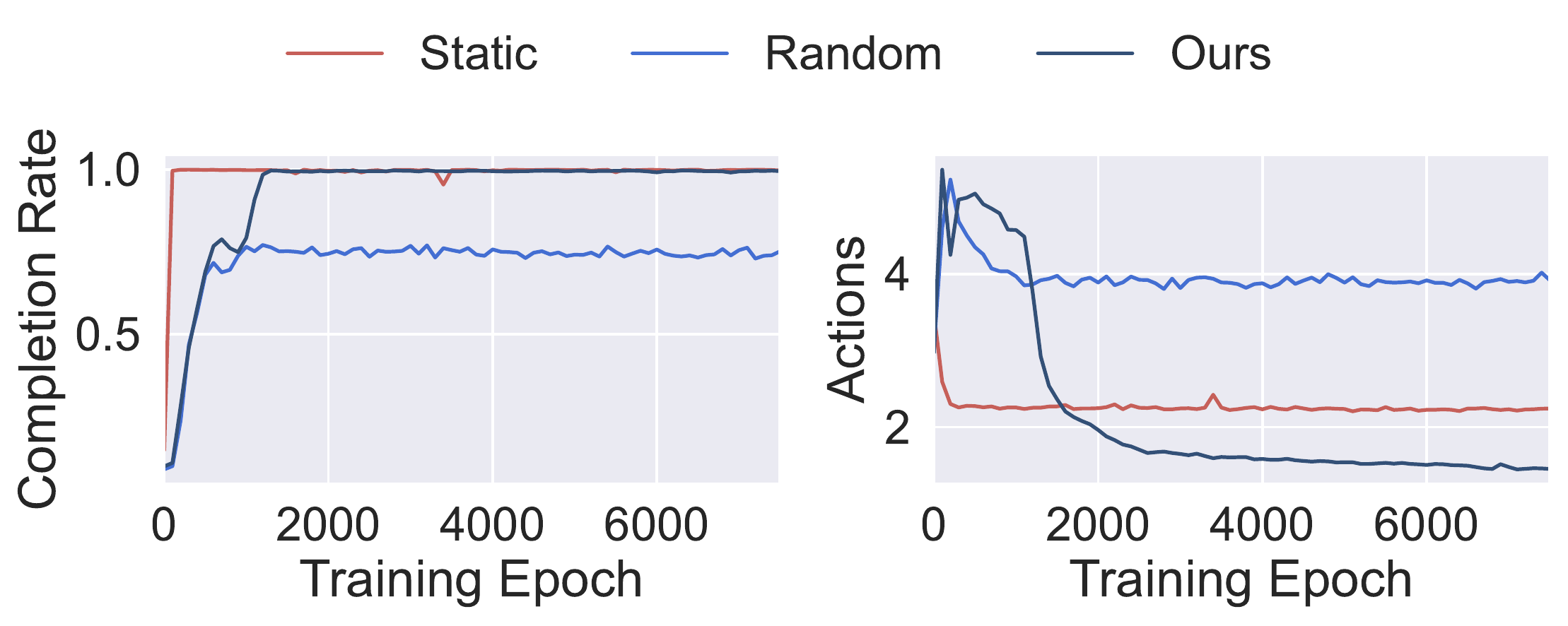}
    \caption{We train our agent till convergence. Left: the fraction of successfully completed episodes per epoch. Ours and Static reach a 100\% successful completion rate. Random does not converge. Right: The number of actions needed on average during a successful episode. Our method needs less actions compared to Static and Random.}
    \Description{Training loss figures with two graphs. Loss graphs of a static interface, a random and the proposed method are presented. The first graph shows the fraction of successfully completed episodes for each method, where the proposed method and the static interface one are the best. The second graph presents average number of actions needed to complete an episode. The proposed method achieheves a better performance than others.}
    \label{fig:training}
\end{figure}

We evaluate the training of our method against a static and a random interface. In the random interface, items are randomly assigned to the slots. 
\Figure{training} shows the \useragent's task completion rate and number of actions per task of all three interfaces during training. Ours and the static baseline converge, whereas the random baseline does not. Furthermore, the mean number of actions of ours is lower than the mean of the static interface.

\paragraph{Generalization to unseen goals} 
\begin{figure}
    \centering
    \vspace{0.8cm}
    \includegraphics[width=0.4\columnwidth]{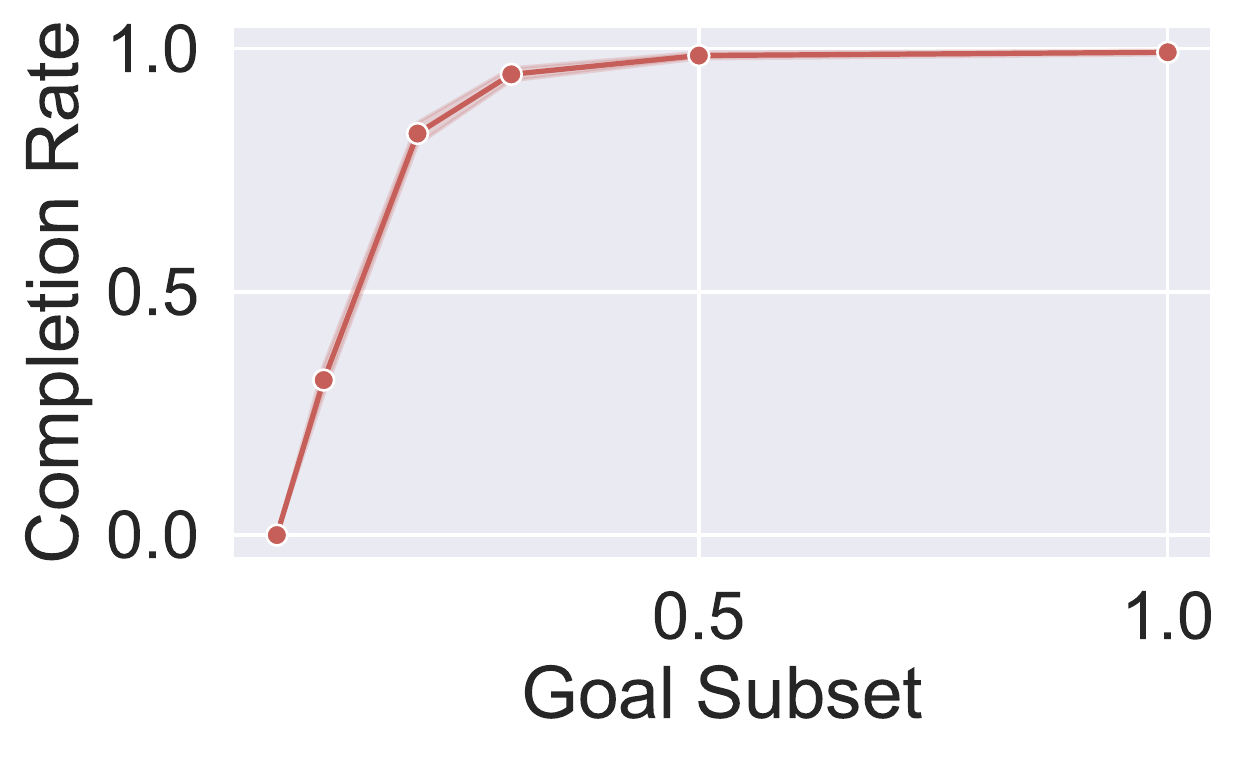}
    \caption{
    The fraction of successfully completed episodes as function of the fraction of the total number of goals. The graph shows that it is sufficient to see half of the goals to learn policies that generalize to all goals.}
    \Description{A graph shows the relation between task completion rate and the fraction of goals the proposed method needs to access. It is showed that the task completion rate increases until top and stays stable after the fraction of goals reach half of the goal subset.}
    \label{fig:goal}
\end{figure}
To understand how well our approach can generalize to unseen goals, we ablate the fraction of goals the agents have access to during training. We then evaluate the learned policies against the full set of goals, which is defined as all possible combinations of character attributes. 
The results are presented in \Figure{goal}. We find that having access to roughly half of the goals is sufficient to not impact the results. This indicates that our approach generalizes to unseen goals of the same set. 

\paragraph{Understanding policy behavior}
\begin{figure}
    \centering
    \includegraphics[width=\columnwidth]{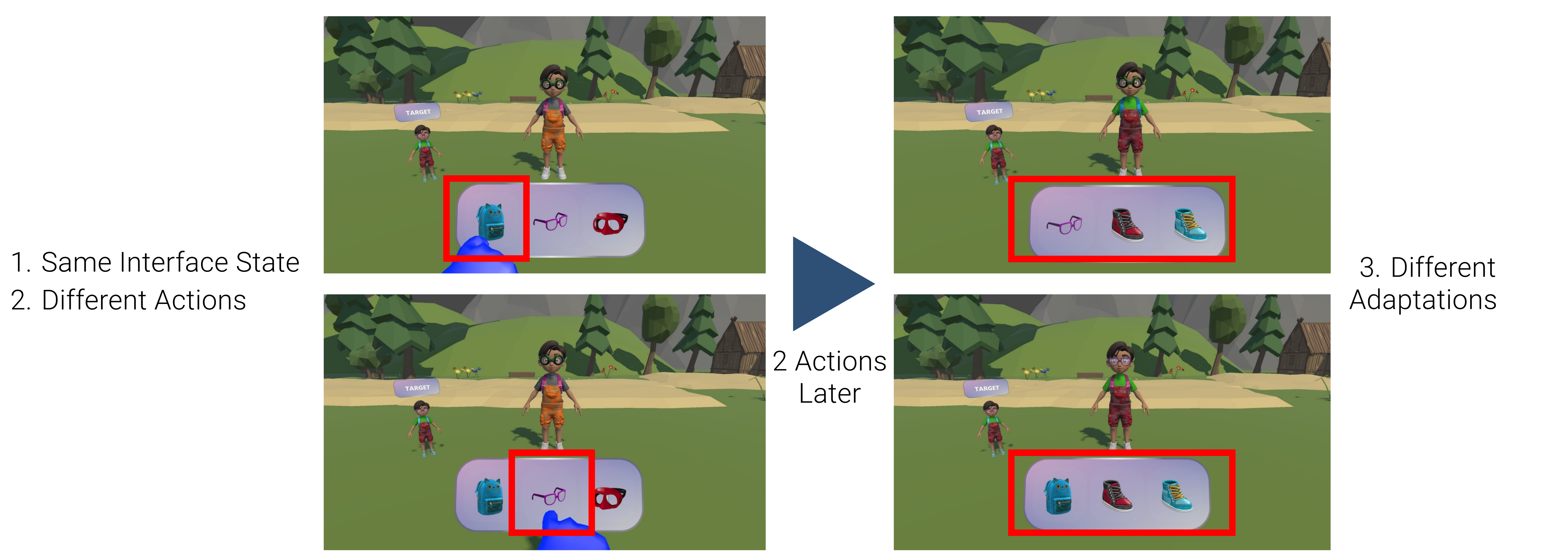}
    \caption{With our method, multiple relevant items can be assigned simultaneously; yet the user can only select one (left). Depending on the user's action (top: select backpack, bottom: select glasses), other item gets assigned later (top: shoes, bottom: backpack). This shows that our method actively adapts to user input.}
    \label{fig:sequence}
    \Description{Four figures are presented to show how the interface change its actions based on the actions of the user agent. The top left figure shows that the user agent selects a backpack and top right figure shows how the interface agent adapts the interface accordingly, by proposing new items. Bottom left figure shows that the user agent glasses and this time the interface agent adapts the interface in a different way, since the action of the user agent is different.}
\end{figure}

We qualitatively analyzed the learned policies of our \interfaceagent to understand how it supports the \useragent in its task.
In \Figure{sequence}, we show a snapshot of two sequences with identical initialization. 
To reach the target character configuration, the user agent can either select the blue bag or the purple glasses (both are needed). 
Depending on which item the user selects at this time step, the \interfaceagent proposes different suggestions in subsequent steps. For instance, the blue backpack the user did not select initially (\Figure{sequence}, bottom) gets suggested again later. 
This behavior shows that the \interfaceagent implicitly reasons about the attribute that the user intends to select based on previous interactions. \addiui{In short, the \interfaceagent learns to suggest items that the user is not wearing, or that the user has not interacted with.}

\subsection{User Study}
\label{sec:userstudy}
Our goal was to create a \useragent whose behavior resembles that of real users, so the \interfaceagent can support them in the same task. To this end, we evaluated the sim-to-real transfer capabilities of our framework by conducting a user study where the \interfaceagent interacted with participants instead of the \useragent.

\subsubsection{Baselines}
We compared our method to two supervised learning methods and the static interface (see \Sec{task}).
In line with previous work \cite{gebhardt2019learning}, we used a Support Vector Machine (SVM) with a Radial basis function (RBF) kernel as a baseline. We used the implementations of scikit-learn \cite{scikit-learn} and optimized the hyperparameters for performance. The feature vector of the baseline was identical to that of our method. The baseline learned the probability with which a user will select a certain character attribute next. We assigned the three attributes with the highest probability to the menu slots. Note that we did not consider "Next" to be an item. 

\paragraph{Dataset} We collected data from 6 participants to train the supervised baselines. These participants did not take part in the user study. They interacted with the static interface, which resulted in a dataset with over 3000 logged interactions. We found that more data points did not improve the performance of the SVM classifier through k-fold cross-validation and reached around 91\% \addiui{top-3 classification accuracy (ie., the percentage of how often the users' selected item was in the top three of the SVM output)} on a test set. Furthermore, we found that the baseline generalize well to unseen participants (again through cross-validation). 

\paragraph{Metrics}
We used two metrics (dependent variables) to evaluate our approach. 
\begin{enumerate}
    \item \emph{Number of Action:} the number of clicks a user needed to complete a task, which is a direct measure of user efficiency \cite{card1980klm}. 
    \item \emph{Task Completion Time:} the total time a user needed to complete a task.
\end{enumerate}

\subsubsection{Procedure}
Participants interacted with the \interfaceagent and the two baselines. The three settings were counterbalanced with a Latin square design, and the participants completed 30 trials per setting. In each condition, we discarded the first six trials for training. The participants were instructed to solve the task as fast as possible while reducing the number of redundant actions. They were allowed to rest in-between trials. We ensured that the number of initial attribute differences between the target and current character was uniformly distributed within the participant's trials. Participants used an Oculus Quest 2 with its controller. 

We recruited 12 participants from staff and students of an institution of higher education (10 male, 2 female, aged between 23 and 33). All participants were right-handed and had a normal or correct-to-normal vision. On average, they needed between 35 to 40 minutes to complete the study.

\subsubsection{Results}
\begin{figure}
     \centering
     \begin{subfigure}[b]{0.3\columnwidth}
        \centering
        \includegraphics[width=\textwidth]{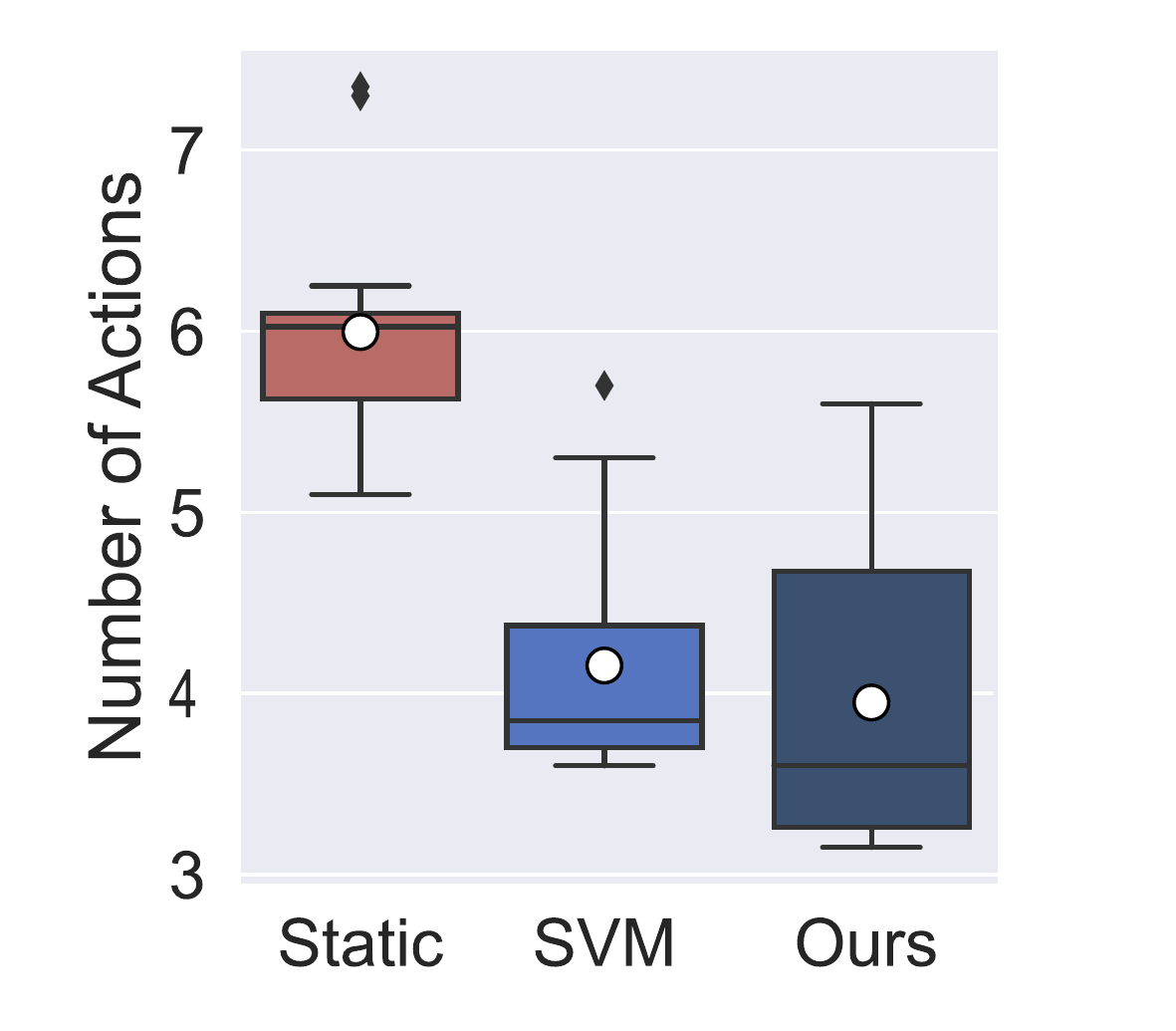}
        \label{fig:userstudy_actions}
    \end{subfigure}
    \begin{subfigure}{0.3\columnwidth}
        \centering
        \includegraphics[width=\textwidth]{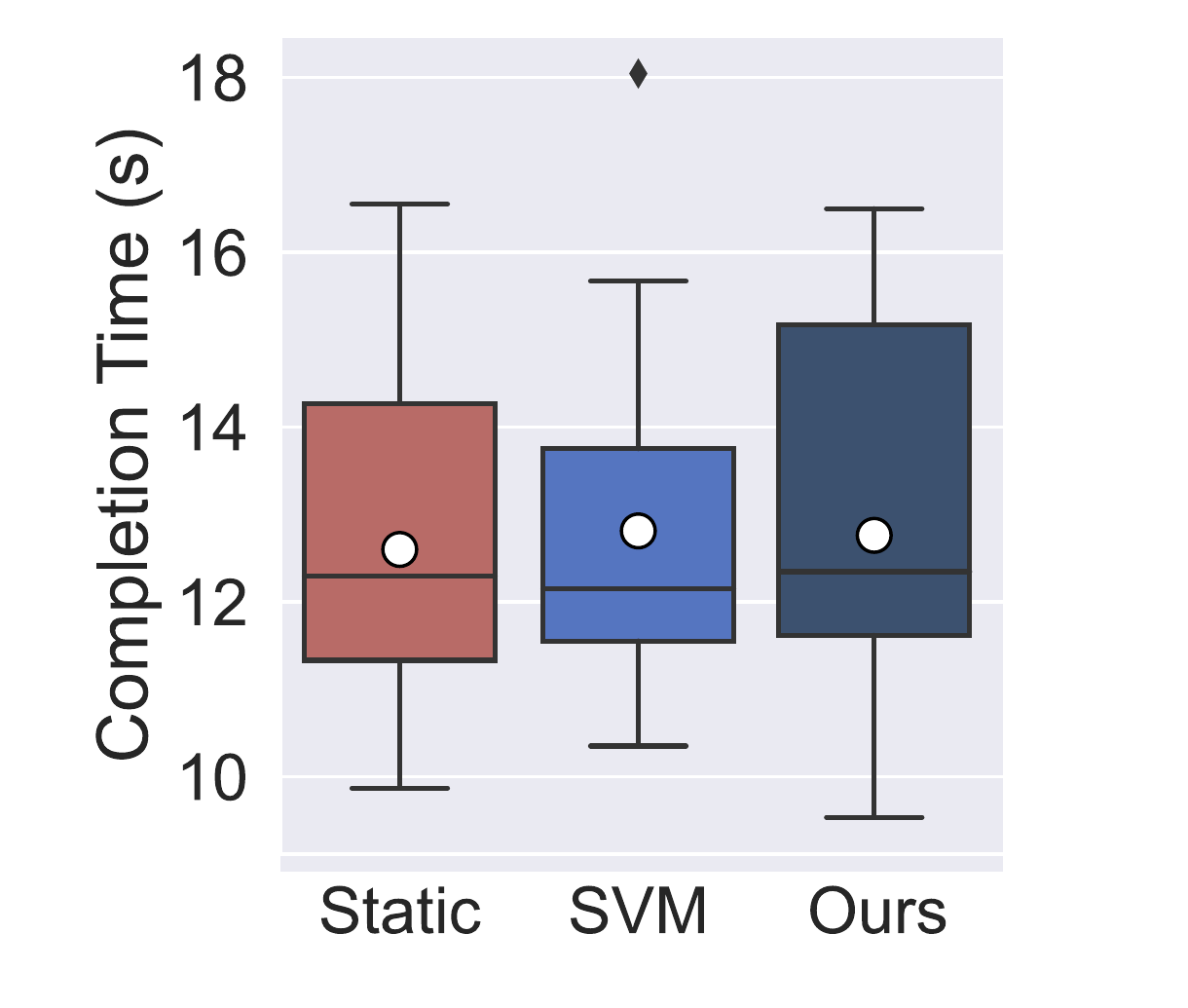}
        \label{fig:userstudy_time}
    \end{subfigure}
    \caption{The average number of actions (left) and the task completion time (right) participants needed to finish the tasks of our user study. We compare Ours against Static, an SVM, and a Bayes classifier. Our approach performs similar to the two data-driven approaches and uses significantly fewer actions than the static baseline. There is no significant effect on the task completion time.}
    \Description{
        Two bar graphs are presented to show user study results. The first bar plot shows the number of actions from 3 to 7 on the Y-axis against the four bars presented for four different methods: 1. Static, 2. SVM, 3. Bayes and 4. Ours, i.e., the proposed method. The proposed method and the Bayes are the best ones in terms of number of actions. The second bar plot shows the completion time from 10s to 18s on the Y-axis against four bars for four different methods again. Again, the proposed method and the Bayes are the best methods in terms of task completion time. 
    }
    \label{fig:userstudy}
\end{figure}

We present a summary of our results in \Figure{userstudy}. We analyzed the effect of conditions on the performance of participants with respect to the number of actions and task completion time. 

Participants needed on average $4.23 \pm 1.20$ actions to complete a task with our method, compared to $6.19 \pm 0.89$, and $4.40 \pm 0.78$ for the static, and SVM baselines respectively. To analyze the data, we performed a Friedman test as normality was violated (Shapiro-Wilk). We found a significant effect of the method on the number of actions $(\chi^2(2) = 39.94, p<.001)$. Conover's posthoc test revealed significant differences between the static interface and the two other methods (all $p<0.001$). We also found an significant effect between our method and the SVM $p=0.020$.

When looking at the task completion time, participants using our method needed $12.76 \pm 2.49$ seconds to complete the task. The completion time was $12.6 \pm 2.25$ for the static interface and $12.8 \pm 2.38$. The task completion time was normally distributed (Shapiro-Wilk $p>0.05$). We found no significant difference in overall task completion time with a Greenhouse-Geisser (for sphericity) corrected repeated-measures ANOVA ($F(1.36, 14.96)=1.70, p=0.22$).  

\subsection{Discussion}
To analyze if our multi-agent method is competitive with a baseline that requires carefully collected real user data, we compared it against a supervised SVM. We did not find significant differences between the two methods in the task performance metrics of the number of actions and task completion time. This suggests that our approach is a competitive alternative to data-driven methods for creating adaptive user interfaces. 

The adaptive methods significantly reduce the number of actions necessary to complete the task compared to the static interface. However, no significant differences in task completion times were found. We argue that this could be due to real users being more familiar with the ordering of items in the static interface that is kept constant across trials. This familiarity is not captured by our current cognitive model or incentivized in the reward function. In future work, we will model familiarity and investigate its effect on task completion time. 

We have shown qualitatively that our \interfaceagent learns to take previous user actions into account. This characteristic is core to meaningful adaptations. At the moment, our agent's capabilities are limited by the size of the stack $\stack$. In the future, recurrent methods such as LSTM could be investigated to overcome this limitation. 

Furthermore, we presented evidence that our method can generalize to goals that were not seen during training. 
It is important to mention that the results of this experiment are subject to its task and that seen and unseen goals are from the same distribution. Nevertheless, the study provides first indications that our approach generalizes to real-world applications where users' goals might not always be encountered during training.
\section{Additional Usecases}
We introduce three additional use cases in VR to demonstrate how our approach generalizes to different scenarios. Note that our method only requires minimal adaptations across tasks. Please refer to our supplementary video for visual demonstrations of the tasks. Because of the different nature of the tasks, we will report either number of clicks or task completion time. In \Appendix{hierarchicalmenu}, we demonstrate in another use case that our method can also support users in 2D settings.

\subsection*{Number Entry}
\begin{figure}
    \centering
    \includegraphics[width=0.6\columnwidth]{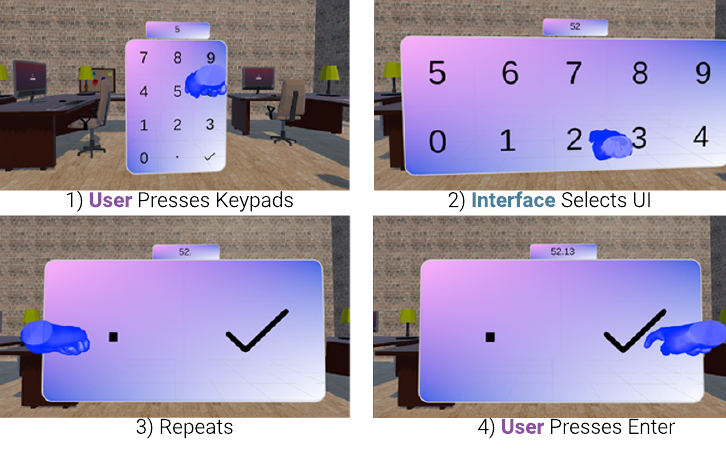}
    \caption{Adaptive keypad: the user agent is asked to enter a randomly initialized price by using a keypad (1). The interface agent can select from three different widgets (2-3): i) a normal keypad, ii) a digits-only keypad, and iii) a non-digits-only keypad. The user agent selects a button of the chosen widget. The task ends when the user agent presses enter (4).}
    \Description{
        A series of five figures is presented that describes each step for adaptive keypad. A keypad is presented in each figure. The first figure describes the initialization. The keypad is initialized as a standard keypad that consists of all numbers. The second, third and fourth figures shows how the interface agent changes the keypad based on the previous keys pressed by the user agent. The fifth and last figure describes that the user finally completed the task and pressed the enter. 
    }
    \label{fig:price}
\end{figure}
We introduce a price entry task on an adaptive keypad to show that our approach can support applications requiring users to issue command sequences and provide meaningful help given users' progress in the task (see \Fig{price}). 
The task assumes a setting where the simulated user must enter a product price between $10.00$ and $99.99$.
To complete the task, the user agent has to enter the first two digits, the decimal point, the second two digits, and then press enter. 
The interface agent can select one of three different interface layouts: i) a standard keypad, ii) a keypad with only digits and iii) a widget with only the decimal point and the enter key (see \Fig{price}.2). 

The goal difference penalty (\Eq{rd}) in this case is based on whether the current price $\tools$ matches the target price $\gattr$: 
\begin{equation}
    \error_{gd} = -\sum_t\mathbbm{1}_{\tools_t \neq \gattr_t},
\end{equation}
where $\mathbbm{1}$ is an indicator that is 1 if $\tools_t \neq \gattr_t$ and $0$ otherwise, and $t$ is the current timestep. Every time a button is hit, $t$ increases by 1. This is similar to the penalty in all other tasks. However, it considers that the order of the entries matters. This task converges in 20 hours and 1500 iterations. On average, the \useragent needs $4.0$ seconds to complete the task in cooperation with the \interfaceagent, compared to $4.9$ seconds when using a static keypad. \add{The number of clicks is identical, since the full task can be solved on the standard keypad.}  

\addiui{
\subsubsection*{Qualitative Policy Inspection} We observe that the \interfaceagent learns to select the UI that has the biggest buttons for an expected number entry. From this we can conclude that the \interfaceagent implicitly learns the concept of Speed-Accuracy trade off.  
}

\subsection*{Block Building}
\label{sec:building}
\begin{figure}
    \centering
    \includegraphics[width=0.6\columnwidth]{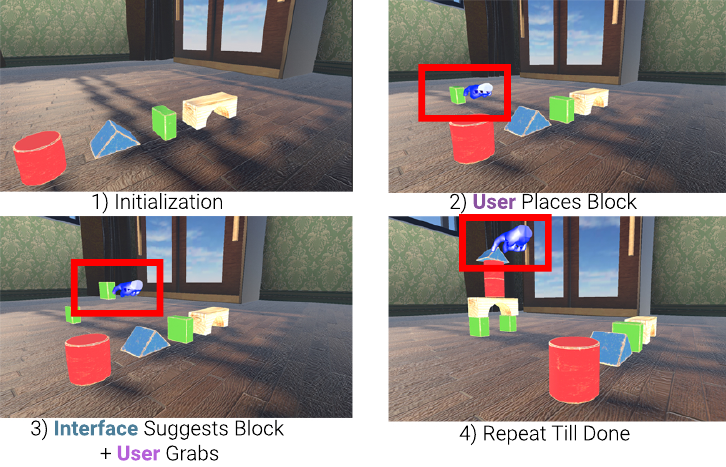}
    \vspace{-3mm}
    \caption{Block Building: The \useragent is building a castle from blocks (1). The \useragent places the first block (2). The \interfaceagent suggests a next block to place (3). This is repeated till the castle is built (4). }
    \Description{
        A series of four figures is presented that describes each step for block building task. Different building blocks with different colors are presented in each figure. The target brush is green circle. The first figure describes the initialization. The second figure shows that the user agent selects and places a block. The third figure shows that the interface agent suggests a block to user agent and user agent grabs it. The fourth figure shows that this process is continued until the task is completed.
    }
    \label{fig:building}
\end{figure}

The second scenario is a block-building task (\Fig{building}) where the \useragent constructs various castle-like structures from blocks. It can choose between 4 blocks (wall, gate, tower, roof) and a delete button. The agent needs to move the hand to a staging place for the blocks (see \Figure{building}) and then place the block in the corresponding location. The block cannot be placed in the air, i.e., it always needs another block on the floor below. The \interfaceagent suggests a next block every time the \useragent places a block. However, the \useragent can put the block down, in case it is unsuitable. An action is picking or placing a block. 

This task represents a subset of tasks that do not have a Heads-Up-Display-like UI to interact with, but are situated directly in the virtual world. This is a common interactive experience of AR/VR systems. This task takes 3000 iterations for both agents to converge. The \useragent needs on average $1.1$ actions with our method, compared to $2.0$ actions without the \interfaceagent. Thus 1.1 indicates that the \interfaceagent suggests the correct next block, most of the time. 

\addiui{
\subsubsection*{Qualitative Policy Inspection} We observe that the policy learns to always suggest a block that is usable given the current state of the tower. This indicates that the policy has an implicit understanding of the order of blocks and can distinguish between those belonging to the foundation versus the upper parts of a tower.
}

\subsection*{Out-of-reach Item Grabbing}
\label{sec:out-of-reach}

\begin{figure}
    \centering
    \includegraphics[width=0.6\columnwidth]{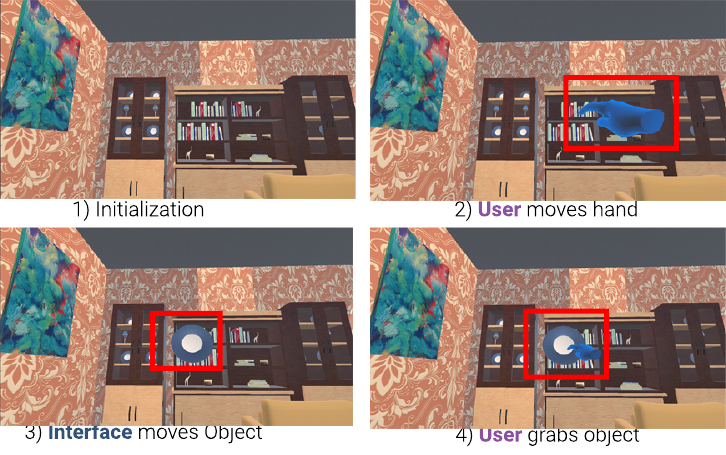}
    \caption{Out-of-reach object grabbing: the user agent attempts to grab a specific object, that is initially out of reach, in a space containing multiple objects (1). The user agent learned to move towards an object to indicate its intention to grab it (2). Based on that, the interface agent learned to move the intended object within the user agent's reach (3). The user agent then grabs the object to finish the task (4).}
    \Description{
        A series of four figures is presented that describes out-of-reach object grabbing task. The first figure describes the initialization, where there is a bookshelf and there are two cupboards, one which has plates inside it and the other one is empty. The second figure shows that the user agent moves its hand towards the cupboard filled with plates. The third figure shows how the interface agent reacts and removes a plate from the cupboard and bring it forward. The fourth and the last figure shows that the user agent grabs the object brought forward by the interface agent and completes the task.
    }
    \label{fig:objectgrab}
\end{figure}

In the third usage scenario, the user needs to grab an object that is initially out of reach. Thus, the \interfaceagent needs to move an object within reach of the \useragent, which can then grab it. The \interfaceagent observes the location of the users end-effector. The task environment includes several objects such that \useragent and \interfaceagent need to collaborate to select the correct target object and complete the task (see \Figure{objectgrab}). This scenario represents tasks that users cannot solve without the help of an adaptive component. Such a setting commonly arises in UIs of emerging technologies \cite{gebhardt2021optimal}.
Our method needed 150 iterations to converge and training ran ca. an hour. 

In this use case, we changed the lower level of our \useragent to learn motor control with RL instead of using the Fitts-Law-based motor controller. This highlights the modularity of our approach and can be useful in scenarios where existing models, such as Fitt's Law, are not sufficient. We introduce the reinforcement-learning-based motor controller in \Appendix{learnedUser}.

\addiui{
\subsubsection*{Qualitative Policy Inspection} We qualitatively evaluate the learned policy. We find that the the policy selects objects positioned in the direction of the users' arm movement, rather than the closest ones. This indicates that the policy implicitly learns about the correlation between directionality and intent. 
}

\addiui{\section{Policy behavior}
We have introduced a method that can be applied to vastly different scenarios, from a toolbar in Virtual Reality, to building blocks in a 3D setting, to handing out of reach items. For all these different scenarios the policy learned different behavior. For the toolbar scenario, the policy learned to suggest items the user has not interacted with before or has not already selected. For the number entry, the \interfaceagent learned an implicit understanding of speed-accuracy trade-off. For block building it learned the concept of ordering suggestions in the right sequence. For out-of-reach-item grabbing the policy learned to link intent with directionality.

What all of these components have in common is that the \interfaceagent learned to support the user, without having access to data. We enabled the \interfaceagent to learn the underlying task structure, by treating human-computer interaction as multiplayer cooperative game. The wide variety of use cases we have shown, with the same method formulation, is a strong indicator that this is a first step towards methods that are not application specific handcrafted or rely on offline collected user data.
}

\section{Limitations \& Future Work}
\label{sec:limitations}

MARLUI is a novel and general approach that offers exciting possibilities for adaptive user interfaces. It models human-computer interaction as a multiplayer cooperative game by teaching a simulated user agent and an interface agent to cooperate.
Learned policies of the interface agent have shown their capability to effectively assist real users.
The successful demonstrations of our approach in a wide variety of use cases pose a first step towards general methods that are not tied to specific applications nor dependent on manually crafted rules or offline user data collection.
\addiui{However, there are limitations that require further research.}

\addiui{In our evaluation, we have compared against a supervised learning baseline. We did not find any differences in terms of task completion time or of number of actions necessary. This indicated that our data-free method is on par with data driven methods.  This is important, since data can be hard time obtain, unavaible, or of low quality. However, there is room to outperform learned approaches. Since we do not rely on data, we could, for instance, investigate using multiple user models with different levels of skill more easily. Also, investigating continuous learning schemes could enable UIs that develop with the user. Crucially, and similar to our method, this would remain a form of cooperative game.}

From an end-user perspective, we assume that the user knows what they are trying to achieve and reach a specific goal state. Removing the dependency on an explicit goal would open up more application domains, such as opened-ended creativity tasks like drawing. This presents an interesting future research direction, where we could learn to infer the user's intended goal from interactions with the interface. Having an informed goal, would enable a goal-oriented RL, which we use in our paper. To achieve this, we could use inverse RL \cite{ng2000algorithms} to learn a reward function of the \useragent from unlabeled user data, which is cheap to collect. Investigating to what extent these approaches produce reward functions that generalize to unseen user goals remains an open question.

Moreover, user goals can change dynamically during system usage in HCI, particularly in creative tasks where users may have a broad range of objectives. This presents challenges for standard RL approaches, which assume that user goals remain stationary. Future research on MARL for AUIs could focus on finding strategies to easily adapt trained interfaces to new user goals. This would yield more robust and flexible adaptive interfaces.  

We have demonstrated that our formulation can solve problems with up to 5 billion possible states (character creation application \Sec{task}). However, the complexity of the problem grows exponentially with the number of states. This makes it challenging for MARLUI to scale to interfaces with even larger state spaces. To overcome this, we could explore different input modalities, such as representing the state of the UI as an image instead of using a one-hot encoding. This is similar to work on RL agents playing video games \cite{mnih2013playing}, which showed that image representations can be effective ways to deal with large state spaces.

We have shown that the simulated \useragent's behavior was sufficiently human-like to enable the \interfaceagent to learn helpful policies that transfer to real users. The \interfaceagent's performance is inherently limited by the \useragent. Therefore, increasing realism in the model of the simulated user is an interesting future research direction, for instance, achieving human-like gaze patterns \cite{chen2015emergence} or motor control using a biomechanical  model \cite{fischer2021reinforcement}.

Our method has theoretical appeal because it provides a plausible model of the bilateral nature of AUIs: the adaptation depends on the user, whereas also the user action depends on the adaptation. Modeling this unilaterally as in supervised learning does not reflect reality well. Treating Adaptive UIs as MARL enables a better understanding of how users interact with a UI and how AUIs need to be adapted. Our setup has the potential to scale to multiple users with different skills and intentions. This can lead to bespoke assistive UIs for users with specific needs or UIs for users with specific expertise levels. In line with current research \cite{murray2022simulation}, we believe that future work can leverage our method to gain a better theoretical understanding of how users interact with a UI. 

In summary, MARLUI is a promising approach that opens up a range of exciting research directions for adaptive UIs that leverage RL. While some limitations exist, such as the complexity challenge and the non-stationarity of user goals, future work can build on our approach to make further strides in this field.
\section{Conclusion}
We have taken a first step towards a general reinforcement learning-based framework for adaptive UIs. We introduced a multi-agent reinforcement learning approach that does not rely on any pre-collected user data or task-specific knowledge\addiui{, while performing on par with data-driven methods.}. Our method features a \useragent and an \interfaceagent. The \useragent tries to achieve a task-dependent goal as fast as possible, while the \interfaceagent learns the underlying task structure by observing the interactions between the \useragent and the UI. Since the \useragent is RL-based and therefore learns through trial-and-error interactions with the interface, it does not require real user data. We have evaluated our approach in simulation and with humans in a variety of different tasks. Results have shown that our method performs on par with data-driven baselines that rely on task and interface-specific data.
This indicates that MARLUI poses a first step towards general methods for adaptive interfaces that are not tied to specific applications nor dependent on offline user data collection.


\begin{acks}
Left empty for review. 
\end{acks}

\bibliographystyle{ACM-Reference-Format}
\bibliography{references}


\begin{thebibliography}{102}


\ifx \showCODEN    \undefined \def \showCODEN     #1{\unskip}     \fi
\ifx \showDOI      \undefined \def \showDOI       #1{#1}\fi
\ifx \showISBNx    \undefined \def \showISBNx     #1{\unskip}     \fi
\ifx \showISBNxiii \undefined \def \showISBNxiii  #1{\unskip}     \fi
\ifx \showISSN     \undefined \def \showISSN      #1{\unskip}     \fi
\ifx \showLCCN     \undefined \def \showLCCN      #1{\unskip}     \fi
\ifx \shownote     \undefined \def \shownote      #1{#1}          \fi
\ifx \showarticletitle \undefined \def \showarticletitle #1{#1}   \fi
\ifx \showURL      \undefined \def \showURL       {\relax}        \fi
\providecommand\bibfield[2]{#2}
\providecommand\bibinfo[2]{#2}
\providecommand\natexlab[1]{#1}
\providecommand\showeprint[2][]{arXiv:#2}

\bibitem[Anderson et~al\mbox{.}(1997)]%
        {anderson1997act}
\bibfield{author}{\bibinfo{person}{John~R Anderson}, \bibinfo{person}{Michael
  Matessa}, {and} \bibinfo{person}{Christian Lebiere}.}
  \bibinfo{year}{1997}\natexlab{}.
\newblock \showarticletitle{ACT-R: A theory of higher level cognition and its
  relation to visual attention}.
\newblock \bibinfo{journal}{\emph{Human--Computer Interaction}}
  \bibinfo{volume}{12}, \bibinfo{number}{4} (\bibinfo{year}{1997}),
  \bibinfo{pages}{439--462}.
\newblock


\bibitem[{\AA}str{\"o}m(1965)]%
        {aastrom1965optimal}
\bibfield{author}{\bibinfo{person}{Karl~Johan {\AA}str{\"o}m}.}
  \bibinfo{year}{1965}\natexlab{}.
\newblock \showarticletitle{Optimal control of Markov processes with incomplete
  state information}.
\newblock \bibinfo{journal}{\emph{Journal of mathematical analysis and
  applications}} \bibinfo{volume}{10}, \bibinfo{number}{1}
  (\bibinfo{year}{1965}), \bibinfo{pages}{174--205}.
\newblock


\bibitem[Bailly et~al\mbox{.}(2016)]%
        {Bailly2017}
\bibfield{author}{\bibinfo{person}{Gilles Bailly}, \bibinfo{person}{Eric
  Lecolinet}, {and} \bibinfo{person}{Laurence Nigay}.}
  \bibinfo{year}{2016}\natexlab{}.
\newblock \showarticletitle{Visual Menu Techniques}.
\newblock \bibinfo{journal}{\emph{ACM Comput. Surv.}} \bibinfo{volume}{49},
  \bibinfo{number}{4}, Article \bibinfo{articleno}{60} (\bibinfo{date}{dec}
  \bibinfo{year}{2016}), \bibinfo{numpages}{41}~pages.
\newblock
\showISSN{0360-0300}
\urldef\tempurl%
\url{https://doi.org/10.1145/3002171}
\showDOI{\tempurl}


\bibitem[Baker et~al\mbox{.}(2019)]%
        {baker2019emergent}
\bibfield{author}{\bibinfo{person}{Bowen Baker}, \bibinfo{person}{Ingmar
  Kanitscheider}, \bibinfo{person}{Todor Markov}, \bibinfo{person}{Yi Wu},
  \bibinfo{person}{Glenn Powell}, \bibinfo{person}{Bob McGrew}, {and}
  \bibinfo{person}{Igor Mordatch}.} \bibinfo{year}{2019}\natexlab{}.
\newblock \showarticletitle{Emergent tool use from multi-agent autocurricula}.
\newblock \bibinfo{journal}{\emph{arXiv preprint arXiv:1909.07528}}
  (\bibinfo{year}{2019}).
\newblock


\bibitem[Berry et~al\mbox{.}(2011)]%
        {Berry2011}
\bibfield{author}{\bibinfo{person}{Pauline~M Berry}, \bibinfo{person}{Melinda
  Gervasio}, \bibinfo{person}{Bart Peintner}, {and} \bibinfo{person}{Neil
  Yorke-Smith}.} \bibinfo{year}{2011}\natexlab{}.
\newblock \showarticletitle{PTIME: Personalized assistance for calendaring}.
\newblock \bibinfo{journal}{\emph{ACM Transactions on Intelligent Systems and
  Technology (TIST)}} \bibinfo{volume}{2}, \bibinfo{number}{4}
  (\bibinfo{year}{2011}), \bibinfo{pages}{1--22}.
\newblock


\bibitem[Bosma and Andr\'{e}(2004)]%
        {Bosma2004}
\bibfield{author}{\bibinfo{person}{Wauter Bosma} {and}
  \bibinfo{person}{Elisabeth Andr\'{e}}.} \bibinfo{year}{2004}\natexlab{}.
\newblock \showarticletitle{Exploiting Emotions to Disambiguate Dialogue Acts}.
  In \bibinfo{booktitle}{\emph{Proceedings of the 9th International Conference
  on Intelligent User Interfaces}} (Funchal, Madeira, Portugal)
  \emph{(\bibinfo{series}{IUI ’04})}. \bibinfo{publisher}{Association for
  Computing Machinery}, \bibinfo{address}{New York, NY, USA},
  \bibinfo{pages}{85–92}.
\newblock
\showISBNx{1581138156}
\urldef\tempurl%
\url{https://doi.org/10.1145/964442.964459}
\showDOI{\tempurl}


\bibitem[Botvinick(2012)]%
        {botvinick2012hierarchical}
\bibfield{author}{\bibinfo{person}{Matthew~Michael Botvinick}.}
  \bibinfo{year}{2012}\natexlab{}.
\newblock \showarticletitle{Hierarchical reinforcement learning and decision
  making}.
\newblock \bibinfo{journal}{\emph{Current opinion in neurobiology}}
  \bibinfo{volume}{22}, \bibinfo{number}{6} (\bibinfo{year}{2012}),
  \bibinfo{pages}{956--962}.
\newblock


\bibitem[Brockman et~al\mbox{.}(2016)]%
        {brockman2016openai}
\bibfield{author}{\bibinfo{person}{Greg Brockman}, \bibinfo{person}{Vicki
  Cheung}, \bibinfo{person}{Ludwig Pettersson}, \bibinfo{person}{Jonas
  Schneider}, \bibinfo{person}{John Schulman}, \bibinfo{person}{Jie Tang},
  {and} \bibinfo{person}{Wojciech Zaremba}.} \bibinfo{year}{2016}\natexlab{}.
\newblock \bibinfo{title}{OpenAI Gym}.
\newblock
\newblock
\showeprint[arxiv]{1606.01540}~[cs.LG]


\bibitem[Browne et~al\mbox{.}(2016)]%
        {Browne1990}
\bibfield{author}{\bibinfo{person}{Dermot Browne}, \bibinfo{person}{Peter
  Totterdell}, {and} \bibinfo{person}{Mike Norman}.}
  \bibinfo{year}{2016}\natexlab{}.
\newblock \bibinfo{booktitle}{\emph{Adaptive user interfaces}}.
\newblock \bibinfo{publisher}{Elsevier}.
\newblock


\bibitem[Card et~al\mbox{.}(1983)]%
        {card1983the}
\bibfield{author}{\bibinfo{person}{S Card}, \bibinfo{person}{T Moran}, {and}
  \bibinfo{person}{A Newell}.} \bibinfo{year}{1983}\natexlab{}.
\newblock \bibinfo{title}{T he Psychology of Human Computer Interaction}.
\newblock
\newblock


\bibitem[Card et~al\mbox{.}(1986)]%
        {card1986model}
\bibfield{author}{\bibinfo{person}{Stuartk Card}, \bibinfo{person}{THOMASP
  MORAN}, {and} \bibinfo{person}{Allen Newell}.}
  \bibinfo{year}{1986}\natexlab{}.
\newblock \showarticletitle{The model human processor- An engineering model of
  human performance}.
\newblock \bibinfo{journal}{\emph{Handbook of perception and human
  performance.}} \bibinfo{volume}{2}, \bibinfo{number}{45--1}
  (\bibinfo{year}{1986}).
\newblock


\bibitem[Card et~al\mbox{.}(1980)]%
        {card1980klm}
\bibfield{author}{\bibinfo{person}{Stuart~K. Card}, \bibinfo{person}{Thomas~P.
  Moran}, {and} \bibinfo{person}{Allen Newell}.}
  \bibinfo{year}{1980}\natexlab{}.
\newblock \showarticletitle{The Keystroke-Level Model for User Performance Time
  with Interactive Systems}.
\newblock \bibinfo{journal}{\emph{Commun. ACM}} \bibinfo{volume}{23},
  \bibinfo{number}{7} (\bibinfo{date}{jul} \bibinfo{year}{1980}),
  \bibinfo{pages}{396–410}.
\newblock
\showISSN{0001-0782}
\urldef\tempurl%
\url{https://doi.org/10.1145/358886.358895}
\showDOI{\tempurl}


\bibitem[Cheema et~al\mbox{.}(2020)]%
        {cheema2020predicting}
\bibfield{author}{\bibinfo{person}{Noshaba Cheema}, \bibinfo{person}{Laura~A
  Frey-Law}, \bibinfo{person}{Kourosh Naderi}, \bibinfo{person}{Jaakko
  Lehtinen}, \bibinfo{person}{Philipp Slusallek}, {and} \bibinfo{person}{Perttu
  H{\"a}m{\"a}l{\"a}inen}.} \bibinfo{year}{2020}\natexlab{}.
\newblock \showarticletitle{Predicting mid-air interaction movements and
  fatigue using deep reinforcement learning}. In
  \bibinfo{booktitle}{\emph{Proceedings of the 2020 CHI Conference on Human
  Factors in Computing Systems}}. \bibinfo{pages}{1--13}.
\newblock


\bibitem[Chen et~al\mbox{.}(2019)]%
        {Chen2019}
\bibfield{author}{\bibinfo{person}{Minmin Chen}, \bibinfo{person}{Alex Beutel},
  \bibinfo{person}{Paul Covington}, \bibinfo{person}{Sagar Jain},
  \bibinfo{person}{Francois Belletti}, {and} \bibinfo{person}{Ed~H Chi}.}
  \bibinfo{year}{2019}\natexlab{}.
\newblock \showarticletitle{Top-K Off-Policy Correction for a REINFORCE
  Recommender System}. In \bibinfo{booktitle}{\emph{Proceedings of the Twelfth
  ACM International Conference on Web Search and Data Mining}}
  \emph{(\bibinfo{series}{WSDM '19})}. ACM, \bibinfo{pages}{456--464}.
\newblock
\urldef\tempurl%
\url{https://doi.org/10.1145/3289600.3290999}
\showDOI{\tempurl}


\bibitem[Chen et~al\mbox{.}(2015)]%
        {chen2015emergence}
\bibfield{author}{\bibinfo{person}{Xiuli Chen}, \bibinfo{person}{Gilles
  Bailly}, \bibinfo{person}{Duncan~P Brumby}, \bibinfo{person}{Antti
  Oulasvirta}, {and} \bibinfo{person}{Andrew Howes}.}
  \bibinfo{year}{2015}\natexlab{}.
\newblock \showarticletitle{The emergence of interactive behavior: A model of
  rational menu search}. In \bibinfo{booktitle}{\emph{Proceedings of the 33rd
  annual ACM conference on human factors in computing systems}}.
  \bibinfo{pages}{4217--4226}.
\newblock


\bibitem[{Christen} et~al\mbox{.}(2021)]%
        {christen2021hide}
\bibfield{author}{\bibinfo{person}{Sammy {Christen}}, \bibinfo{person}{Lukas
  {Jendele}}, \bibinfo{person}{Emre {Aksan}}, {and} \bibinfo{person}{Otmar
  {Hilliges}}.} \bibinfo{year}{2021}\natexlab{}.
\newblock \showarticletitle{Learning Functionally Decomposed Hierarchies for
  Continuous Control Tasks With Path Planning}.
\newblock \bibinfo{journal}{\emph{IEEE Robotics and Automation Letters}}
  \bibinfo{volume}{6}, \bibinfo{number}{2} (\bibinfo{year}{2021}),
  \bibinfo{pages}{3623--3630}.
\newblock
\urldef\tempurl%
\url{https://doi.org/10.1109/LRA.2021.3060403}
\showDOI{\tempurl}


\bibitem[Cockburn et~al\mbox{.}(2007a)]%
        {Cockburn2007}
\bibfield{author}{\bibinfo{person}{Andy Cockburn}, \bibinfo{person}{Carl
  Gutwin}, {and} \bibinfo{person}{Saul Greenberg}.}
  \bibinfo{year}{2007}\natexlab{a}.
\newblock \showarticletitle{A Predictive Model of Menu Performance}. In
  \bibinfo{booktitle}{\emph{Proceedings of the SIGCHI Conference on Human
  Factors in Computing Systems}} (San Jose, California, USA)
  \emph{(\bibinfo{series}{CHI '07})}. \bibinfo{publisher}{Association for
  Computing Machinery}, \bibinfo{address}{New York, NY, USA},
  \bibinfo{pages}{627–636}.
\newblock
\showISBNx{9781595935939}
\urldef\tempurl%
\url{https://doi.org/10.1145/1240624.1240723}
\showDOI{\tempurl}


\bibitem[Cockburn et~al\mbox{.}(2007b)]%
        {10.1145/1240624.1240723}
\bibfield{author}{\bibinfo{person}{Andy Cockburn}, \bibinfo{person}{Carl
  Gutwin}, {and} \bibinfo{person}{Saul Greenberg}.}
  \bibinfo{year}{2007}\natexlab{b}.
\newblock \showarticletitle{A Predictive Model of Menu Performance}. In
  \bibinfo{booktitle}{\emph{Proceedings of the SIGCHI Conference on Human
  Factors in Computing Systems}} (San Jose, California, USA)
  \emph{(\bibinfo{series}{CHI '07})}. \bibinfo{publisher}{Association for
  Computing Machinery}, \bibinfo{address}{New York, NY, USA},
  \bibinfo{pages}{627–636}.
\newblock
\showISBNx{9781595935939}
\urldef\tempurl%
\url{https://doi.org/10.1145/1240624.1240723}
\showDOI{\tempurl}


\bibitem[Dai et~al\mbox{.}(2013)]%
        {PengCrowdsourcing2013}
\bibfield{author}{\bibinfo{person}{Peng Dai}, \bibinfo{person}{Christopher
  Lin}, \bibinfo{person}{Mausam Mausam}, {and} \bibinfo{person}{Daniel Weld}.}
  \bibinfo{year}{2013}\natexlab{}.
\newblock \showarticletitle{POMDP-based control of workflows for
  crowdsourcing}.
\newblock \bibinfo{journal}{\emph{Artificial Intelligence}}
  \bibinfo{volume}{202} (\bibinfo{date}{09} \bibinfo{year}{2013}),
  \bibinfo{pages}{52–85}.
\newblock
\urldef\tempurl%
\url{https://doi.org/10.1016/j.artint.2013.06.002}
\showDOI{\tempurl}


\bibitem[Debard et~al\mbox{.}(2020)]%
        {debard2020multiagent}
\bibfield{author}{\bibinfo{person}{Quentin Debard},
  \bibinfo{person}{Jilles~Steeve Dibangoye}, \bibinfo{person}{St{\'e}phane
  Canu}, {and} \bibinfo{person}{Christian Wolf}.}
  \bibinfo{year}{2020}\natexlab{}.
\newblock \showarticletitle{Learning 3D Navigation Protocols on Touch
  Interfaces with Cooperative Multi-agent Reinforcement Learning}. In
  \bibinfo{booktitle}{\emph{Machine Learning and Knowledge Discovery in
  Databases}}, \bibfield{editor}{\bibinfo{person}{Ulf Brefeld},
  \bibinfo{person}{Elisa Fromont}, \bibinfo{person}{Andreas Hotho},
  \bibinfo{person}{Arno Knobbe}, \bibinfo{person}{Marloes Maathuis}, {and}
  \bibinfo{person}{C{\'e}line Robardet}} (Eds.). \bibinfo{publisher}{Springer
  International Publishing}, \bibinfo{address}{Cham}, \bibinfo{pages}{35--52}.
\newblock
\showISBNx{978-3-030-46133-1}


\bibitem[Denison et~al\mbox{.}(2013)]%
        {denison2013rational}
\bibfield{author}{\bibinfo{person}{Stephanie Denison},
  \bibinfo{person}{Elizabeth Bonawitz}, \bibinfo{person}{Alison Gopnik}, {and}
  \bibinfo{person}{Thomas~L Griffiths}.} \bibinfo{year}{2013}\natexlab{}.
\newblock \showarticletitle{Rational variability in children’s causal
  inferences: The sampling hypothesis}.
\newblock \bibinfo{journal}{\emph{Cognition}} \bibinfo{volume}{126},
  \bibinfo{number}{2} (\bibinfo{year}{2013}), \bibinfo{pages}{285--300}.
\newblock


\bibitem[Duchowski et~al\mbox{.}(2018)]%
        {duchowski2018index}
\bibfield{author}{\bibinfo{person}{Andrew~T Duchowski},
  \bibinfo{person}{Krzysztof Krejtz}, \bibinfo{person}{Izabela Krejtz},
  \bibinfo{person}{Cezary Biele}, \bibinfo{person}{Anna Niedzielska},
  \bibinfo{person}{Peter Kiefer}, \bibinfo{person}{Martin Raubal}, {and}
  \bibinfo{person}{Ioannis Giannopoulos}.} \bibinfo{year}{2018}\natexlab{}.
\newblock \showarticletitle{The index of pupillary activity: Measuring
  cognitive load vis-{\`a}-vis task difficulty with pupil oscillation}. In
  \bibinfo{booktitle}{\emph{Proceedings of the 2018 CHI conference on human
  factors in computing systems}}. \bibinfo{pages}{1--13}.
\newblock


\bibitem[Faulring et~al\mbox{.}(2010)]%
        {Faulring2010}
\bibfield{author}{\bibinfo{person}{Andrew Faulring}, \bibinfo{person}{Brad
  Myers}, \bibinfo{person}{Ken Mohnkern}, \bibinfo{person}{Bradley Schmerl},
  \bibinfo{person}{Aaron Steinfeld}, \bibinfo{person}{John Zimmerman},
  \bibinfo{person}{Asim Smailagic}, \bibinfo{person}{Jeffery Hansen}, {and}
  \bibinfo{person}{Daniel Siewiorek}.} \bibinfo{year}{2010}\natexlab{}.
\newblock \showarticletitle{Agent-Assisted Task Management That Reduces Email
  Overload}. In \bibinfo{booktitle}{\emph{Proceedings of the 15th International
  Conference on Intelligent User Interfaces}} (Hong Kong, China)
  \emph{(\bibinfo{series}{IUI ’10})}. \bibinfo{publisher}{Association for
  Computing Machinery}, \bibinfo{address}{New York, NY, USA},
  \bibinfo{pages}{61–70}.
\newblock
\showISBNx{9781605585154}
\urldef\tempurl%
\url{https://doi.org/10.1145/1719970.1719980}
\showDOI{\tempurl}


\bibitem[Ferretti et~al\mbox{.}(2014)]%
        {ferretti2014exploiting}
\bibfield{author}{\bibinfo{person}{Stefano Ferretti}, \bibinfo{person}{Silvia
  Mirri}, \bibinfo{person}{Catia Prandi}, {and} \bibinfo{person}{Paola
  Salomoni}.} \bibinfo{year}{2014}\natexlab{}.
\newblock \showarticletitle{Exploiting reinforcement learning to profile users
  and personalize web pages}. In \bibinfo{booktitle}{\emph{2014 IEEE 38th
  International Computer Software and Applications Conference Workshops}}.
  IEEE, \bibinfo{pages}{252--257}.
\newblock


\bibitem[Findlater et~al\mbox{.}(2009)]%
        {Findlater2009}
\bibfield{author}{\bibinfo{person}{Leah Findlater}, \bibinfo{person}{Karyn
  Moffatt}, \bibinfo{person}{Joanna McGrenere}, {and} \bibinfo{person}{Jessica
  Dawson}.} \bibinfo{year}{2009}\natexlab{}.
\newblock \showarticletitle{Ephemeral Adaptation: The Use of Gradual Onset to
  Improve Menu Selection Performance}. In \bibinfo{booktitle}{\emph{Proceedings
  of the SIGCHI Conference on Human Factors in Computing Systems}} (Boston, MA,
  USA) \emph{(\bibinfo{series}{CHI '09})}. \bibinfo{publisher}{Association for
  Computing Machinery}, \bibinfo{address}{New York, NY, USA},
  \bibinfo{pages}{1655–1664}.
\newblock
\showISBNx{9781605582467}
\urldef\tempurl%
\url{https://doi.org/10.1145/1518701.1518956}
\showDOI{\tempurl}


\bibitem[Fischer et~al\mbox{.}(2021)]%
        {fischer2021reinforcement}
\bibfield{author}{\bibinfo{person}{Florian Fischer}, \bibinfo{person}{Miroslav
  Bachinski}, \bibinfo{person}{Markus Klar}, \bibinfo{person}{Arthur Fleig},
  {and} \bibinfo{person}{J{\"o}rg M{\"u}ller}.}
  \bibinfo{year}{2021}\natexlab{}.
\newblock \showarticletitle{Reinforcement learning control of a biomechanical
  model of the upper extremity}.
\newblock \bibinfo{journal}{\emph{Scientific Reports}} \bibinfo{volume}{11},
  \bibinfo{number}{1} (\bibinfo{year}{2021}), \bibinfo{pages}{1--15}.
\newblock


\bibitem[Fitts(1954)]%
        {fitts1954information}
\bibfield{author}{\bibinfo{person}{Paul~M Fitts}.}
  \bibinfo{year}{1954}\natexlab{}.
\newblock \showarticletitle{The information capacity of the human motor system
  in controlling the amplitude of movement.}
\newblock \bibinfo{journal}{\emph{Journal of experimental psychology}}
  \bibinfo{volume}{47}, \bibinfo{number}{6} (\bibinfo{year}{1954}),
  \bibinfo{pages}{381}.
\newblock


\bibitem[Foerster et~al\mbox{.}(2016)]%
        {foerster2016learning}
\bibfield{author}{\bibinfo{person}{Jakob Foerster},
  \bibinfo{person}{Ioannis~Alexandros Assael}, \bibinfo{person}{Nando
  De~Freitas}, {and} \bibinfo{person}{Shimon Whiteson}.}
  \bibinfo{year}{2016}\natexlab{}.
\newblock \showarticletitle{Learning to communicate with deep multi-agent
  reinforcement learning}.
\newblock \bibinfo{journal}{\emph{Advances in neural information processing
  systems}}  \bibinfo{volume}{29} (\bibinfo{year}{2016}).
\newblock


\bibitem[Frank and Badre(2012)]%
        {frank2012mechanisms}
\bibfield{author}{\bibinfo{person}{Michael~J Frank} {and}
  \bibinfo{person}{David Badre}.} \bibinfo{year}{2012}\natexlab{}.
\newblock \showarticletitle{Mechanisms of hierarchical reinforcement learning
  in corticostriatal circuits 1: computational analysis}.
\newblock \bibinfo{journal}{\emph{Cerebral cortex}} \bibinfo{volume}{22},
  \bibinfo{number}{3} (\bibinfo{year}{2012}), \bibinfo{pages}{509--526}.
\newblock


\bibitem[Ga{\v{s}}i{\'{c}} and Young(2014)]%
        {Gasic2014}
\bibfield{author}{\bibinfo{person}{Milica Ga{\v{s}}i{\'{c}}} {and}
  \bibinfo{person}{Steve Young}.} \bibinfo{year}{2014}\natexlab{}.
\newblock \showarticletitle{{Gaussian processes for POMDP-based dialogue
  manager optimization}}.
\newblock \bibinfo{journal}{\emph{IEEE Transactions on Audio, Speech and
  Language Processing}} \bibinfo{volume}{22}, \bibinfo{number}{1}
  (\bibinfo{year}{2014}), \bibinfo{pages}{28--40}.
\newblock
\showISSN{15587916}
\urldef\tempurl%
\url{https://doi.org/10.1109/TASL.2013.2282190}
\showDOI{\tempurl}


\bibitem[Gebhardt et~al\mbox{.}(2019)]%
        {gebhardt2019learning}
\bibfield{author}{\bibinfo{person}{Christoph Gebhardt}, \bibinfo{person}{Brian
  Hecox}, \bibinfo{person}{Bas van Opheusden}, \bibinfo{person}{Daniel Wigdor},
  \bibinfo{person}{James Hillis}, \bibinfo{person}{Otmar Hilliges}, {and}
  \bibinfo{person}{Hrvoje Benko}.} \bibinfo{year}{2019}\natexlab{}.
\newblock \showarticletitle{Learning Cooperative Personalized Policies from
  Gaze Data}. In \bibinfo{booktitle}{\emph{Proceedings of the 32nd Annual ACM
  Symposium on User Interface Software and Technology}} (New Orleans, LA, USA)
  \emph{(\bibinfo{series}{UIST '19})}. \bibinfo{publisher}{ACM},
  \bibinfo{address}{New York, NY, USA}, \bibinfo{numpages}{10}~pages.
\newblock
\showISBNx{978-1-4503-6816-2/19/10}
\urldef\tempurl%
\url{https://doi.org/10.1145/3332165.3347933}
\showDOI{\tempurl}


\bibitem[Gebhardt and Hilliges(2021)]%
        {gebhardt2021optimal}
\bibfield{author}{\bibinfo{person}{Christoph Gebhardt} {and}
  \bibinfo{person}{Otmar Hilliges}.} \bibinfo{year}{2021}\natexlab{}.
\newblock \showarticletitle{Optimal Control to Support High-Level User Goals in
  Human-Computer Interaction}.
\newblock In \bibinfo{booktitle}{\emph{Artificial Intelligence for Human
  Computer Interaction: A Modern Approach}}. \bibinfo{publisher}{Springer},
  \bibinfo{pages}{33--72}.
\newblock


\bibitem[Gebhardt et~al\mbox{.}(2021)]%
        {gebhardt2020hierarchical}
\bibfield{author}{\bibinfo{person}{Christoph Gebhardt}, \bibinfo{person}{Antti
  Oulasvirta}, {and} \bibinfo{person}{Otmar Hilliges}.}
  \bibinfo{year}{2021}\natexlab{}.
\newblock \showarticletitle{Hierarchical Reinforcement Learning as a Model of
  Human Task Interleaving}.
\newblock \bibinfo{journal}{\emph{Computational Brain and Behavior}}
  (\bibinfo{year}{2021}).
\newblock
\urldef\tempurl%
\url{https://arxiv.org/pdf/2001.02122.pdf}
\showURL{%
\tempurl}


\bibitem[Gershman et~al\mbox{.}(2015)]%
        {gershman2015computational}
\bibfield{author}{\bibinfo{person}{Samuel~J Gershman}, \bibinfo{person}{Eric~J
  Horvitz}, {and} \bibinfo{person}{Joshua~B Tenenbaum}.}
  \bibinfo{year}{2015}\natexlab{}.
\newblock \showarticletitle{Computational rationality: A converging paradigm
  for intelligence in brains, minds, and machines}.
\newblock \bibinfo{journal}{\emph{Science}} \bibinfo{volume}{349},
  \bibinfo{number}{6245} (\bibinfo{year}{2015}), \bibinfo{pages}{273--278}.
\newblock


\bibitem[Gershman et~al\mbox{.}(2012)]%
        {gershman2012multistability}
\bibfield{author}{\bibinfo{person}{Samuel~J Gershman}, \bibinfo{person}{Edward
  Vul}, {and} \bibinfo{person}{Joshua~B Tenenbaum}.}
  \bibinfo{year}{2012}\natexlab{}.
\newblock \showarticletitle{Multistability and perceptual inference}.
\newblock \bibinfo{journal}{\emph{Neural computation}} \bibinfo{volume}{24},
  \bibinfo{number}{1} (\bibinfo{year}{2012}), \bibinfo{pages}{1--24}.
\newblock


\bibitem[Glowacka(2019)]%
        {glowacka2019bandit}
\bibfield{author}{\bibinfo{person}{Dorota Glowacka}.}
  \bibinfo{year}{2019}\natexlab{}.
\newblock \showarticletitle{Bandit algorithms in recommender systems}. In
  \bibinfo{booktitle}{\emph{Proceedings of the 13th ACM Conference on
  Recommender Systems}}. \bibinfo{pages}{574--575}.
\newblock


\bibitem[Guiard and Rioul(2015)]%
        {guiard2015mathematical}
\bibfield{author}{\bibinfo{person}{Yves Guiard} {and} \bibinfo{person}{Olivier
  Rioul}.} \bibinfo{year}{2015}\natexlab{}.
\newblock \showarticletitle{A mathematical description of the speed/accuracy
  trade-off of aimed movement}. In \bibinfo{booktitle}{\emph{Proceedings of the
  2015 British HCI Conference}}. \bibinfo{pages}{91--100}.
\newblock


\bibitem[Hick(1952)]%
        {hick1952rate}
\bibfield{author}{\bibinfo{person}{William~E Hick}.}
  \bibinfo{year}{1952}\natexlab{}.
\newblock \showarticletitle{On the rate of gain of information}.
\newblock \bibinfo{journal}{\emph{Quarterly Journal of experimental
  psychology}} \bibinfo{volume}{4}, \bibinfo{number}{1} (\bibinfo{year}{1952}),
  \bibinfo{pages}{11--26}.
\newblock


\bibitem[Horvitz et~al\mbox{.}(1998)]%
        {Horvitz1998}
\bibfield{author}{\bibinfo{person}{Eric Horvitz}, \bibinfo{person}{Jack
  Breese}, \bibinfo{person}{David Heckerman}, \bibinfo{person}{David Hovel},
  {and} \bibinfo{person}{Koos Rommelse}.} \bibinfo{year}{1998}\natexlab{}.
\newblock \showarticletitle{The Lumi\`{e}Re Project: Bayesian User Modeling for
  Inferring the Goals and Needs of Software Users}. In
  \bibinfo{booktitle}{\emph{Proceedings of the Fourteenth Conference on
  Uncertainty in Artificial Intelligence}} (Madison, Wisconsin)
  \emph{(\bibinfo{series}{UAI’98})}. \bibinfo{publisher}{Morgan Kaufmann
  Publishers Inc.}, \bibinfo{address}{San Francisco, CA, USA},
  \bibinfo{pages}{256–265}.
\newblock
\showISBNx{155860555X}


\bibitem[Howard(1960)]%
        {howard1960dynamic}
\bibfield{author}{\bibinfo{person}{Ronald~A Howard}.}
  \bibinfo{year}{1960}\natexlab{}.
\newblock \showarticletitle{Dynamic programming and markov processes.}
\newblock  (\bibinfo{year}{1960}).
\newblock


\bibitem[Howes et~al\mbox{.}(2018)]%
        {howes2018interaction}
\bibfield{author}{\bibinfo{person}{Andrew Howes}, \bibinfo{person}{Xiuli Chen},
  \bibinfo{person}{Aditya Acharya}, {and} \bibinfo{person}{Richard~L Lewis}.}
  \bibinfo{year}{2018}\natexlab{}.
\newblock \showarticletitle{Interaction as an emergent property of a Partially
  Observable Markov Decision Process}.
\newblock \bibinfo{journal}{\emph{Computational interaction}}
  (\bibinfo{year}{2018}), \bibinfo{pages}{287--310}.
\newblock


\bibitem[Hu et~al\mbox{.}(2018)]%
        {Hu2018}
\bibfield{author}{\bibinfo{person}{Zehong Hu}, \bibinfo{person}{Yitao Liang},
  \bibinfo{person}{Jie Zhang}, \bibinfo{person}{Zhao Li}, {and}
  \bibinfo{person}{Yang Liu}.} \bibinfo{year}{2018}\natexlab{}.
\newblock \showarticletitle{Inference aided reinforcement learning for
  incentive mechanism design in crowdsourcing}. In
  \bibinfo{booktitle}{\emph{Advances in Neural Information Processing Systems}}
  \emph{(\bibinfo{series}{NIPS '18})}. \bibinfo{pages}{5508--5518}.
\newblock
\urldef\tempurl%
\url{https://arxiv.org/abs/1806.00206}
\showURL{%
\tempurl}


\bibitem[Ismail and Sariff(2018)]%
        {sariff2018multiagent}
\bibfield{author}{\bibinfo{person}{Zool~Hilmi Ismail} {and}
  \bibinfo{person}{Nohaidda Sariff}.} \bibinfo{year}{2018}\natexlab{}.
\newblock \showarticletitle{A Survey and Analysis of Cooperative Multi-Agent
  Robot Systems: Challenges and Directions}.
\newblock In \bibinfo{booktitle}{\emph{Applications of Mobile Robots}},
  \bibfield{editor}{\bibinfo{person}{Efren~Gorrostieta Hurtado}} (Ed.).
  \bibinfo{publisher}{IntechOpen}, \bibinfo{address}{Rijeka}, Chapter~1.
\newblock
\urldef\tempurl%
\url{https://doi.org/10.5772/intechopen.79337}
\showDOI{\tempurl}


\bibitem[Jaderberg et~al\mbox{.}(2019)]%
        {jaderberg2019quake}
\bibfield{author}{\bibinfo{person}{Max Jaderberg}, \bibinfo{person}{Wojciech~M.
  Czarnecki}, \bibinfo{person}{Iain Dunning}, \bibinfo{person}{Luke Marris},
  \bibinfo{person}{Guy Lever}, \bibinfo{person}{Antonio~Garcia Castañeda},
  \bibinfo{person}{Charles Beattie}, \bibinfo{person}{Neil~C. Rabinowitz},
  \bibinfo{person}{Ari~S. Morcos}, \bibinfo{person}{Avraham Ruderman},
  \bibinfo{person}{Nicolas Sonnerat}, \bibinfo{person}{Tim Green},
  \bibinfo{person}{Louise Deason}, \bibinfo{person}{Joel~Z. Leibo},
  \bibinfo{person}{David Silver}, \bibinfo{person}{Demis Hassabis},
  \bibinfo{person}{Koray Kavukcuoglu}, {and} \bibinfo{person}{Thore Graepel}.}
  \bibinfo{year}{2019}\natexlab{}.
\newblock \showarticletitle{Human-level performance in 3D multiplayer games
  with population-based reinforcement learning}.
\newblock \bibinfo{journal}{\emph{Science}} \bibinfo{volume}{364},
  \bibinfo{number}{6443} (\bibinfo{year}{2019}), \bibinfo{pages}{859--865}.
\newblock
\urldef\tempurl%
\url{https://doi.org/10.1126/science.aau6249}
\showDOI{\tempurl}


\bibitem[Jokinen et~al\mbox{.}(2021a)]%
        {jokinen2021touchscreen}
\bibfield{author}{\bibinfo{person}{Jussi Jokinen}, \bibinfo{person}{Aditya
  Acharya}, \bibinfo{person}{Mohammad Uzair}, \bibinfo{person}{Xinhui Jiang},
  {and} \bibinfo{person}{Antti Oulasvirta}.} \bibinfo{year}{2021}\natexlab{a}.
\newblock \showarticletitle{Touchscreen Typing as Optimal Supervisory Control}.
  In \bibinfo{booktitle}{\emph{Proceedings of the 2021 CHI Conference on Human
  Factors in Computing Systems (CHI '21)}}. \bibinfo{publisher}{ACM}.
\newblock
\urldef\tempurl%
\url{https://userinterfaces.aalto.fi/touchscreen-typing/}
\showURL{%
\tempurl}


\bibitem[Jokinen et~al\mbox{.}(2021b)]%
        {jokinen2021multitasking}
\bibfield{author}{\bibinfo{person}{Jussi~PP Jokinen}, \bibinfo{person}{Tuomo
  Kujala}, {and} \bibinfo{person}{Antti Oulasvirta}.}
  \bibinfo{year}{2021}\natexlab{b}.
\newblock \showarticletitle{Multitasking in driving as optimal adaptation under
  uncertainty}.
\newblock \bibinfo{journal}{\emph{Human factors}} \bibinfo{volume}{63},
  \bibinfo{number}{8} (\bibinfo{year}{2021}), \bibinfo{pages}{1324--1341}.
\newblock


\bibitem[Kangas et~al\mbox{.}(2022)]%
        {kangas2022scalable}
\bibfield{author}{\bibinfo{person}{Ioannis Kangas}, \bibinfo{person}{Maud
  Schwoerer}, {and} \bibinfo{person}{Lucas Bernardi}.}
  \bibinfo{year}{2022}\natexlab{}.
\newblock \showarticletitle{Scalable User Interface Optimization Using
  Combinatorial Bandits}. In \bibinfo{booktitle}{\emph{Proceedings of the 45th
  International ACM SIGIR Conference on Research and Development in Information
  Retrieval}}. \bibinfo{pages}{3375--3379}.
\newblock


\bibitem[Kieras and Meyer(1997)]%
        {kieras1997overview}
\bibfield{author}{\bibinfo{person}{Davis~E Kieras} {and}
  \bibinfo{person}{Davis~E Meyer}.} \bibinfo{year}{1997}\natexlab{}.
\newblock \showarticletitle{An overview of the EPIC architecture for cognition
  and performance with application to human-computer interaction}.
\newblock \bibinfo{journal}{\emph{Human--Computer Interaction}}
  \bibinfo{volume}{12}, \bibinfo{number}{4} (\bibinfo{year}{1997}),
  \bibinfo{pages}{391--438}.
\newblock


\bibitem[Koch et~al\mbox{.}(2019)]%
        {koch2019may}
\bibfield{author}{\bibinfo{person}{Janin Koch}, \bibinfo{person}{Andr{\'e}s
  Lucero}, \bibinfo{person}{Lena Hegemann}, {and} \bibinfo{person}{Antti
  Oulasvirta}.} \bibinfo{year}{2019}\natexlab{}.
\newblock \showarticletitle{May AI? Design ideation with cooperative contextual
  bandits}. In \bibinfo{booktitle}{\emph{Proceedings of the 2019 CHI Conference
  on Human Factors in Computing Systems}}. \bibinfo{pages}{1--12}.
\newblock


\bibitem[Kolekar et~al\mbox{.}(2010)]%
        {kolekar2010learning}
\bibfield{author}{\bibinfo{person}{Sucheta~V Kolekar},
  \bibinfo{person}{Sriram~G Sanjeevi}, {and} \bibinfo{person}{DS Bormane}.}
  \bibinfo{year}{2010}\natexlab{}.
\newblock \showarticletitle{Learning style recognition using artificial neural
  network for adaptive user interface in e-learning}. In
  \bibinfo{booktitle}{\emph{2010 IEEE International conference on computational
  intelligence and computing research}}. IEEE, \bibinfo{pages}{1--5}.
\newblock


\bibitem[Koyama et~al\mbox{.}(2014)]%
        {Koyama2014}
\bibfield{author}{\bibinfo{person}{Yuki Koyama}, \bibinfo{person}{Daisuke
  Sakamoto}, {and} \bibinfo{person}{Takeo Igarashi}.}
  \bibinfo{year}{2014}\natexlab{}.
\newblock \showarticletitle{{Crowd-powered parameter analysis for visual design
  exploration}}.
\newblock \bibinfo{journal}{\emph{Proceedings of the 27th annual ACM symposium
  on User interface software and technology - UIST '14}}
  (\bibinfo{year}{2014}), \bibinfo{pages}{65--74}.
\newblock
\showISBNx{9781450330695}
\urldef\tempurl%
\url{https://doi.org/10.1145/2642918.2647386}
\showDOI{\tempurl}


\bibitem[Koyama et~al\mbox{.}(2016)]%
        {Koyama2016}
\bibfield{author}{\bibinfo{person}{Yuki Koyama}, \bibinfo{person}{Daisuke
  Sakamoto}, {and} \bibinfo{person}{Takeo Igarashi}.}
  \bibinfo{year}{2016}\natexlab{}.
\newblock \showarticletitle{{SelPh : Progressive Learning and Support of Manual
  Photo Color Enhancement}}.
\newblock \bibinfo{journal}{\emph{Proc. of CHI '16}} (\bibinfo{year}{2016}).
\newblock
\showISBNx{9781450333627}
\urldef\tempurl%
\url{https://doi.org/10.1145/2858036.2858111}
\showDOI{\tempurl}


\bibitem[Lashkari et~al\mbox{.}(1997)]%
        {Lashkari1997}
\bibfield{author}{\bibinfo{person}{Yezdi Lashkari}, \bibinfo{person}{Max
  Metral}, {and} \bibinfo{person}{Pattie Maes}.}
  \bibinfo{year}{1997}\natexlab{}.
\newblock \showarticletitle{Collaborative interface agents}.
\newblock \bibinfo{journal}{\emph{Readings in agents}} (\bibinfo{year}{1997}),
  \bibinfo{pages}{111--116}.
\newblock


\bibitem[Leibo et~al\mbox{.}(2017)]%
        {leibo2017multi}
\bibfield{author}{\bibinfo{person}{Joel~Z Leibo}, \bibinfo{person}{Vinicius
  Zambaldi}, \bibinfo{person}{Marc Lanctot}, \bibinfo{person}{Janusz Marecki},
  {and} \bibinfo{person}{Thore Graepel}.} \bibinfo{year}{2017}\natexlab{}.
\newblock \showarticletitle{Multi-agent reinforcement learning in sequential
  social dilemmas}.
\newblock \bibinfo{journal}{\emph{arXiv preprint arXiv:1702.03037}}
  (\bibinfo{year}{2017}).
\newblock


\bibitem[Leino et~al\mbox{.}(2019)]%
        {leino2019computer}
\bibfield{author}{\bibinfo{person}{Katri Leino}, \bibinfo{person}{Kashyap
  Todi}, \bibinfo{person}{Antti Oulasvirta}, {and} \bibinfo{person}{Mikko
  Kurimo}.} \bibinfo{year}{2019}\natexlab{}.
\newblock \showarticletitle{Computer-Supported Form Design Using
  Keystroke-Level Modeling with Reinforcement Learning}. In
  \bibinfo{booktitle}{\emph{Proceedings of the 24th International Conference on
  Intelligent User Interfaces: Companion}} (Marina del Ray, California)
  \emph{(\bibinfo{series}{IUI '19})}. \bibinfo{publisher}{Association for
  Computing Machinery}, \bibinfo{address}{New York, NY, USA},
  \bibinfo{pages}{85–86}.
\newblock
\showISBNx{9781450366731}
\urldef\tempurl%
\url{https://doi.org/10.1145/3308557.3308704}
\showDOI{\tempurl}


\bibitem[Liang et~al\mbox{.}(2018)]%
        {liang2018rllib}
\bibfield{author}{\bibinfo{person}{Eric Liang}, \bibinfo{person}{Richard Liaw},
  \bibinfo{person}{Philipp Moritz}, \bibinfo{person}{Robert Nishihara},
  \bibinfo{person}{Roy Fox}, \bibinfo{person}{Ken Goldberg},
  \bibinfo{person}{Joseph~E. Gonzalez}, \bibinfo{person}{Michael~I. Jordan},
  {and} \bibinfo{person}{Ion Stoica}.} \bibinfo{year}{2018}\natexlab{}.
\newblock \bibinfo{title}{RLlib: Abstractions for Distributed Reinforcement
  Learning}.
\newblock
\newblock
\showeprint[arxiv]{1712.09381}~[cs.AI]


\bibitem[Liebman et~al\mbox{.}(2015)]%
        {Liebman2015}
\bibfield{author}{\bibinfo{person}{Elad Liebman}, \bibinfo{person}{Maytal
  Saar-Tsechansky}, {and} \bibinfo{person}{Peter Stone}.}
  \bibinfo{year}{2015}\natexlab{}.
\newblock \showarticletitle{DJ-MC: A Reinforcement-Learning Agent for Music
  Playlist Recommendation}. In \bibinfo{booktitle}{\emph{Proceedings of the
  2015 International Conference on Autonomous Agents and Multiagent Systems}}
  \emph{(\bibinfo{series}{AAMAS '15})}. \bibinfo{pages}{591--599}.
\newblock
\urldef\tempurl%
\url{https://arxiv.org/abs/1401.1880}
\showURL{%
\tempurl}


\bibitem[Lindlbauer et~al\mbox{.}(2019)]%
        {lindlbauer2019context}
\bibfield{author}{\bibinfo{person}{David Lindlbauer},
  \bibinfo{person}{Anna~Maria Feit}, {and} \bibinfo{person}{Otmar Hilliges}.}
  \bibinfo{year}{2019}\natexlab{}.
\newblock \showarticletitle{Context-aware online adaptation of mixed reality
  interfaces}. In \bibinfo{booktitle}{\emph{Proceedings of the 32nd annual ACM
  symposium on user interface software and technology}}.
  \bibinfo{pages}{147--160}.
\newblock


\bibitem[Liu et~al\mbox{.}(2018)]%
        {Liu2018}
\bibfield{author}{\bibinfo{person}{Feng Liu}, \bibinfo{person}{Ruiming Tang},
  \bibinfo{person}{Xutao Li}, \bibinfo{person}{Weinan Zhang},
  \bibinfo{person}{Yunming Ye}, \bibinfo{person}{Haokun Chen},
  \bibinfo{person}{Huifeng Guo}, {and} \bibinfo{person}{Yuzhou Zhang}.}
  \bibinfo{year}{2018}\natexlab{}.
\newblock \showarticletitle{Deep reinforcement learning based recommendation
  with explicit user-item interactions modeling}.
\newblock \bibinfo{journal}{\emph{arXiv preprint arXiv:1810.12027}}
  (\bibinfo{year}{2018}).
\newblock
\urldef\tempurl%
\url{https://arxiv.org/abs/1810.12027}
\showURL{%
\tempurl}


\bibitem[Lomas et~al\mbox{.}(2016)]%
        {lomas2016interface}
\bibfield{author}{\bibinfo{person}{J~Derek Lomas}, \bibinfo{person}{Jodi
  Forlizzi}, \bibinfo{person}{Nikhil Poonwala}, \bibinfo{person}{Nirmal Patel},
  \bibinfo{person}{Sharan Shodhan}, \bibinfo{person}{Kishan Patel},
  \bibinfo{person}{Ken Koedinger}, {and} \bibinfo{person}{Emma Brunskill}.}
  \bibinfo{year}{2016}\natexlab{}.
\newblock \showarticletitle{Interface design optimization as a multi-armed
  bandit problem}. In \bibinfo{booktitle}{\emph{Proceedings of the 2016 CHI
  conference on human factors in computing systems}}.
  \bibinfo{pages}{4142--4153}.
\newblock


\bibitem[Long et~al\mbox{.}(2020)]%
        {epciclr2020}
\bibfield{author}{\bibinfo{person}{Qian Long}, \bibinfo{person}{Zihan Zhou},
  \bibinfo{person}{Abhinav Gupta}, \bibinfo{person}{Fei Fang},
  \bibinfo{person}{Yi Wu}, {and} \bibinfo{person}{Xiaolong Wang}.}
  \bibinfo{year}{2020}\natexlab{}.
\newblock \showarticletitle{Evolutionary Population Curriculum for Scaling
  Multi-Agent Reinforcement Learning}. In
  \bibinfo{booktitle}{\emph{International Conference on Learning
  Representations}}.
\newblock


\bibitem[Lowe et~al\mbox{.}(2017)]%
        {lowe2017multi}
\bibfield{author}{\bibinfo{person}{Ryan Lowe}, \bibinfo{person}{Yi~I Wu},
  \bibinfo{person}{Aviv Tamar}, \bibinfo{person}{Jean Harb},
  \bibinfo{person}{OpenAI Pieter~Abbeel}, {and} \bibinfo{person}{Igor
  Mordatch}.} \bibinfo{year}{2017}\natexlab{}.
\newblock \showarticletitle{Multi-agent actor-critic for mixed
  cooperative-competitive environments}.
\newblock \bibinfo{journal}{\emph{Advances in neural information processing
  systems}}  \bibinfo{volume}{30} (\bibinfo{year}{2017}).
\newblock


\bibitem[Mackay(2000)]%
        {mackay2000responding}
\bibfield{author}{\bibinfo{person}{Wendy Mackay}.}
  \bibinfo{year}{2000}\natexlab{}.
\newblock \showarticletitle{Responding to cognitive overload: Co-adaptation
  between users and technology}.
\newblock \bibinfo{journal}{\emph{Intellectica}} \bibinfo{volume}{30},
  \bibinfo{number}{1} (\bibinfo{year}{2000}), \bibinfo{pages}{177--193}.
\newblock


\bibitem[Maes(1995)]%
        {Maes1995}
\bibfield{author}{\bibinfo{person}{Pattie Maes}.}
  \bibinfo{year}{1995}\natexlab{}.
\newblock \showarticletitle{Agents that reduce work and information overload}.
\newblock In \bibinfo{booktitle}{\emph{Readings in human--computer
  interaction}}. \bibinfo{publisher}{Elsevier}, \bibinfo{pages}{811--821}.
\newblock


\bibitem[McCreath et~al\mbox{.}(2006)]%
        {McCreath2006}
\bibfield{author}{\bibinfo{person}{Eric McCreath}, \bibinfo{person}{Judy Kay},
  {and} \bibinfo{person}{Elisabeth Crawford}.} \bibinfo{year}{2006}\natexlab{}.
\newblock \showarticletitle{IEMS-an approach that combines handcrafted rules
  with learnt instance based rules.}
\newblock \bibinfo{journal}{\emph{Aust. J. Intell. Inf. Process. Syst.}}
  \bibinfo{volume}{9}, \bibinfo{number}{1} (\bibinfo{year}{2006}),
  \bibinfo{pages}{40--53}.
\newblock


\bibitem[Mehrotra and Hendley(2015)]%
        {Mehrotra2015}
\bibfield{author}{\bibinfo{person}{Abhinav Mehrotra} {and}
  \bibinfo{person}{Robert Hendley}.} \bibinfo{year}{2015}\natexlab{}.
\newblock \showarticletitle{{Designing Content-driven Intelligent Notification
  Mechanisms for Mobile Applications}}.
\newblock  (\bibinfo{year}{2015}), \bibinfo{pages}{813--824}.
\newblock
\showISBNx{9781450335744}


\bibitem[Mnih et~al\mbox{.}(2013)]%
        {mnih2013playing}
\bibfield{author}{\bibinfo{person}{Volodymyr Mnih}, \bibinfo{person}{Koray
  Kavukcuoglu}, \bibinfo{person}{David Silver}, \bibinfo{person}{Alex Graves},
  \bibinfo{person}{Ioannis Antonoglou}, \bibinfo{person}{Daan Wierstra}, {and}
  \bibinfo{person}{Martin Riedmiller}.} \bibinfo{year}{2013}\natexlab{}.
\newblock \showarticletitle{Playing atari with deep reinforcement learning}.
\newblock \bibinfo{journal}{\emph{arXiv preprint arXiv:1312.5602}}
  (\bibinfo{year}{2013}).
\newblock


\bibitem[Murray-Smith et~al\mbox{.}(2022)]%
        {murray2022simulation}
\bibfield{author}{\bibinfo{person}{Roderick Murray-Smith},
  \bibinfo{person}{Antti Oulasvirta}, \bibinfo{person}{Andrew Howes},
  \bibinfo{person}{J{\"o}rg M{\"u}ller}, \bibinfo{person}{Aleksi Ikkala},
  \bibinfo{person}{Miroslav Bachinski}, \bibinfo{person}{Arthur Fleig},
  \bibinfo{person}{Florian Fischer}, {and} \bibinfo{person}{Markus Klar}.}
  \bibinfo{year}{2022}\natexlab{}.
\newblock \showarticletitle{What simulation can do for HCI research}.
\newblock \bibinfo{journal}{\emph{Interactions}} \bibinfo{volume}{29},
  \bibinfo{number}{6} (\bibinfo{year}{2022}), \bibinfo{pages}{48--53}.
\newblock


\bibitem[Ng et~al\mbox{.}(2000)]%
        {ng2000algorithms}
\bibfield{author}{\bibinfo{person}{Andrew~Y Ng}, \bibinfo{person}{Stuart
  Russell}, {et~al\mbox{.}}} \bibinfo{year}{2000}\natexlab{}.
\newblock \showarticletitle{Algorithms for inverse reinforcement learning.}. In
  \bibinfo{booktitle}{\emph{Icml}}, Vol.~\bibinfo{volume}{1}.
  \bibinfo{pages}{2}.
\newblock


\bibitem[Ota(2006)]%
        {ota2006multiagent}
\bibfield{author}{\bibinfo{person}{Jun Ota}.} \bibinfo{year}{2006}\natexlab{}.
\newblock \showarticletitle{Multi-agent robot systems as distributed autonomous
  systems}.
\newblock \bibinfo{journal}{\emph{Advanced Engineering Informatics}}
  \bibinfo{volume}{20}, \bibinfo{number}{1} (\bibinfo{year}{2006}),
  \bibinfo{pages}{59--70}.
\newblock
\showISSN{1474-0346}
\urldef\tempurl%
\url{https://doi.org/10.1016/j.aei.2005.06.002}
\showDOI{\tempurl}


\bibitem[Oulasvirta et~al\mbox{.}(2020)]%
        {combinatorialoptimizationdesign2020oulasvirta}
\bibfield{author}{\bibinfo{person}{Antti Oulasvirta},
  \bibinfo{person}{Niraj~Ramesh Dayama}, \bibinfo{person}{Morteza Shiripour},
  \bibinfo{person}{Maximilian John}, {and} \bibinfo{person}{Andreas
  Karrenbauer}.} \bibinfo{year}{2020}\natexlab{}.
\newblock \showarticletitle{Combinatorial Optimization of Graphical User
  Interface Designs}.
\newblock \bibinfo{journal}{\emph{Proc. IEEE}} \bibinfo{volume}{108},
  \bibinfo{number}{3} (\bibinfo{year}{2020}), \bibinfo{pages}{434--464}.
\newblock
\urldef\tempurl%
\url{https://doi.org/10.1109/JPROC.2020.2969687}
\showDOI{\tempurl}


\bibitem[Oulasvirta et~al\mbox{.}(2017)]%
        {functionalityselection2017oulasvirta}
\bibfield{author}{\bibinfo{person}{Antti Oulasvirta}, \bibinfo{person}{Anna
  Feit}, \bibinfo{person}{Perttu L\"{a}hteenlahti}, {and}
  \bibinfo{person}{Andreas Karrenbauer}.} \bibinfo{year}{2017}\natexlab{}.
\newblock \showarticletitle{Computational Support for Functionality Selection
  in Interaction Design}.
\newblock  \bibinfo{volume}{24}, \bibinfo{number}{5}, Article
  \bibinfo{articleno}{34} (\bibinfo{date}{oct} \bibinfo{year}{2017}),
  \bibinfo{numpages}{30}~pages.
\newblock
\showISSN{1073-0516}
\urldef\tempurl%
\url{https://doi.org/10.1145/3131608}
\showDOI{\tempurl}


\bibitem[Oulasvirta et~al\mbox{.}(2022)]%
        {oulasvirta2022computational}
\bibfield{author}{\bibinfo{person}{Antti Oulasvirta}, \bibinfo{person}{Jussi~PP
  Jokinen}, {and} \bibinfo{person}{Andrew Howes}.}
  \bibinfo{year}{2022}\natexlab{}.
\newblock \showarticletitle{Computational Rationality as a Theory of
  Interaction}. In \bibinfo{booktitle}{\emph{CHI Conference on Human Factors in
  Computing Systems}}. \bibinfo{pages}{1--14}.
\newblock


\bibitem[Oulasvirta et~al\mbox{.}(2018)]%
        {oulasvirta2018computational}
\bibfield{author}{\bibinfo{person}{Antti Oulasvirta}, \bibinfo{person}{Per~Ola
  Kristensson}, \bibinfo{person}{Xiaojun Bi}, {and} \bibinfo{person}{Andrew
  Howes}.} \bibinfo{year}{2018}\natexlab{}.
\newblock \bibinfo{booktitle}{\emph{Computational interaction}}.
\newblock \bibinfo{publisher}{Oxford University Press}.
\newblock


\bibitem[Park et~al\mbox{.}(2018)]%
        {park2018adam}
\bibfield{author}{\bibinfo{person}{Seonwook Park}, \bibinfo{person}{Christoph
  Gebhardt}, \bibinfo{person}{Roman R{\"a}dle}, \bibinfo{person}{Anna~Maria
  Feit}, \bibinfo{person}{Hana Vrzakova}, \bibinfo{person}{Niraj~Ramesh
  Dayama}, \bibinfo{person}{Hui-Shyong Yeo}, \bibinfo{person}{Clemens~N
  Klokmose}, \bibinfo{person}{Aaron Quigley}, \bibinfo{person}{Antti
  Oulasvirta}, {et~al\mbox{.}}} \bibinfo{year}{2018}\natexlab{}.
\newblock \showarticletitle{Adam: Adapting multi-user interfaces for
  collaborative environments in real-time}. In
  \bibinfo{booktitle}{\emph{Proceedings of the 2018 CHI conference on human
  factors in computing systems}}. \bibinfo{pages}{1--14}.
\newblock


\bibitem[Pedregosa et~al\mbox{.}(2011)]%
        {scikit-learn}
\bibfield{author}{\bibinfo{person}{F. Pedregosa}, \bibinfo{person}{G.
  Varoquaux}, \bibinfo{person}{A. Gramfort}, \bibinfo{person}{V. Michel},
  \bibinfo{person}{B. Thirion}, \bibinfo{person}{O. Grisel},
  \bibinfo{person}{M. Blondel}, \bibinfo{person}{P. Prettenhofer},
  \bibinfo{person}{R. Weiss}, \bibinfo{person}{V. Dubourg}, \bibinfo{person}{J.
  Vanderplas}, \bibinfo{person}{A. Passos}, \bibinfo{person}{D. Cournapeau},
  \bibinfo{person}{M. Brucher}, \bibinfo{person}{M. Perrot}, {and}
  \bibinfo{person}{E. Duchesnay}.} \bibinfo{year}{2011}\natexlab{}.
\newblock \showarticletitle{Scikit-learn: Machine Learning in {P}ython}.
\newblock \bibinfo{journal}{\emph{Journal of Machine Learning Research}}
  \bibinfo{volume}{12} (\bibinfo{year}{2011}), \bibinfo{pages}{2825--2830}.
\newblock


\bibitem[Pejovic and Musolesi(2014)]%
        {Pejovic2014}
\bibfield{author}{\bibinfo{person}{Veljko Pejovic} {and} \bibinfo{person}{Mirco
  Musolesi}.} \bibinfo{year}{2014}\natexlab{}.
\newblock \showarticletitle{{InterruptMe: Designing Intelligent Prompting
  Mechanisms for Pervasive Applications}}.
\newblock \bibinfo{journal}{\emph{Proceedings of the 2014 ACM International
  Joint Conference on Pervasive and Ubiquitous Computing}}
  (\bibinfo{year}{2014}), \bibinfo{pages}{897--908}.
\newblock
\showISBNx{978-1-4503-2968-2}
\urldef\tempurl%
\url{https://doi.org/10.1145/2632048.2632062}
\showDOI{\tempurl}


\bibitem[Polydoros and Nalpantidis(2017)]%
        {polydoros2017survey}
\bibfield{author}{\bibinfo{person}{Athanasios~S Polydoros} {and}
  \bibinfo{person}{Lazaros Nalpantidis}.} \bibinfo{year}{2017}\natexlab{}.
\newblock \showarticletitle{Survey of model-based reinforcement learning:
  Applications on robotics}.
\newblock \bibinfo{journal}{\emph{Journal of Intelligent \& Robotic Systems}}
  \bibinfo{volume}{86}, \bibinfo{number}{2} (\bibinfo{year}{2017}),
  \bibinfo{pages}{153--173}.
\newblock


\bibitem[Rich et~al\mbox{.}(2005)]%
        {Rich2005}
\bibfield{author}{\bibinfo{person}{Charles Rich}, \bibinfo{person}{Candy
  Sidner}, \bibinfo{person}{Neal Lesh}, \bibinfo{person}{Andrew Garland},
  \bibinfo{person}{Shane Booth}, {and} \bibinfo{person}{Markus Chimani}.}
  \bibinfo{year}{2005}\natexlab{}.
\newblock \showarticletitle{DiamondHelp: A collaborative interface framework
  for networked home appliances}. In \bibinfo{booktitle}{\emph{25th IEEE
  International Conference on Distributed Computing Systems Workshops}}. IEEE,
  \bibinfo{pages}{514--519}.
\newblock


\bibitem[Rich and Sidner(1998)]%
        {Rich1998}
\bibfield{author}{\bibinfo{person}{Charles Rich} {and}
  \bibinfo{person}{Candace~L Sidner}.} \bibinfo{year}{1998}\natexlab{}.
\newblock \showarticletitle{COLLAGEN: A collaboration manager for software
  interface agents}.
\newblock In \bibinfo{booktitle}{\emph{Computational Models of Mixed-Initiative
  Interaction}}. \bibinfo{publisher}{Springer}, \bibinfo{pages}{149--184}.
\newblock


\bibitem[Rizzoglio et~al\mbox{.}(2021)]%
        {RIZZOGLIO2021}
\bibfield{author}{\bibinfo{person}{Fabio Rizzoglio}, \bibinfo{person}{Maura
  Casadio}, \bibinfo{person}{Dalia {De Santis}}, {and}
  \bibinfo{person}{Ferdinando~A. Mussa-Ivaldi}.}
  \bibinfo{year}{2021}\natexlab{}.
\newblock \showarticletitle{Building an adaptive interface via unsupervised
  tracking of latent manifolds}.
\newblock \bibinfo{journal}{\emph{Neural Networks}}  \bibinfo{volume}{137}
  (\bibinfo{year}{2021}), \bibinfo{pages}{174--187}.
\newblock
\showISSN{0893-6080}
\urldef\tempurl%
\url{https://doi.org/10.1016/j.neunet.2021.01.009}
\showDOI{\tempurl}


\bibitem[Salvucci(2001)]%
        {salvucci2001integrated}
\bibfield{author}{\bibinfo{person}{Dario~D Salvucci}.}
  \bibinfo{year}{2001}\natexlab{}.
\newblock \showarticletitle{An integrated model of eye movements and visual
  encoding}.
\newblock \bibinfo{journal}{\emph{Cognitive Systems Research}}
  \bibinfo{volume}{1}, \bibinfo{number}{4} (\bibinfo{year}{2001}),
  \bibinfo{pages}{201--220}.
\newblock


\bibitem[Schulman et~al\mbox{.}(2017)]%
        {schulman2017proximal}
\bibfield{author}{\bibinfo{person}{John Schulman}, \bibinfo{person}{Filip
  Wolski}, \bibinfo{person}{Prafulla Dhariwal}, \bibinfo{person}{Alec Radford},
  {and} \bibinfo{person}{Oleg Klimov}.} \bibinfo{year}{2017}\natexlab{}.
\newblock \bibinfo{title}{Proximal Policy Optimization Algorithms}.
\newblock
\newblock
\showeprint[arxiv]{1707.06347}~[cs.LG]


\bibitem[Sears and Shneiderman(1994)]%
        {Sears1994}
\bibfield{author}{\bibinfo{person}{Andrew Sears} {and} \bibinfo{person}{Ben
  Shneiderman}.} \bibinfo{year}{1994}\natexlab{}.
\newblock \showarticletitle{Split Menus: Effectively Using Selection Frequency
  to Organize Menus}.
\newblock \bibinfo{journal}{\emph{ACM Trans. Comput.-Hum. Interact.}}
  \bibinfo{volume}{1}, \bibinfo{number}{1} (\bibinfo{date}{mar}
  \bibinfo{year}{1994}), \bibinfo{pages}{27–51}.
\newblock
\showISSN{1073-0516}
\urldef\tempurl%
\url{https://doi.org/10.1145/174630.174632}
\showDOI{\tempurl}


\bibitem[Seo and Zhang(2000)]%
        {seo2000reinforcement}
\bibfield{author}{\bibinfo{person}{Young-Woo Seo} {and}
  \bibinfo{person}{Byoung-Tak Zhang}.} \bibinfo{year}{2000}\natexlab{}.
\newblock \showarticletitle{A reinforcement learning agent for personalized
  information filtering}. In \bibinfo{booktitle}{\emph{Proceedings of the 5th
  international conference on Intelligent user interfaces}}.
  \bibinfo{pages}{248--251}.
\newblock


\bibitem[Shahriari et~al\mbox{.}(2015)]%
        {shahriari2015taking}
\bibfield{author}{\bibinfo{person}{Bobak Shahriari}, \bibinfo{person}{Kevin
  Swersky}, \bibinfo{person}{Ziyu Wang}, \bibinfo{person}{Ryan~P Adams}, {and}
  \bibinfo{person}{Nando De~Freitas}.} \bibinfo{year}{2015}\natexlab{}.
\newblock \showarticletitle{Taking the human out of the loop: A review of
  Bayesian optimization}.
\newblock \bibinfo{journal}{\emph{Proc. IEEE}} \bibinfo{volume}{104},
  \bibinfo{number}{1} (\bibinfo{year}{2015}), \bibinfo{pages}{148--175}.
\newblock


\bibitem[Shapley(1953)]%
        {shapley1953stochastic}
\bibfield{author}{\bibinfo{person}{Lloyd~S Shapley}.}
  \bibinfo{year}{1953}\natexlab{}.
\newblock \showarticletitle{Stochastic games}.
\newblock \bibinfo{journal}{\emph{Proceedings of the national academy of
  sciences}} \bibinfo{volume}{39}, \bibinfo{number}{10} (\bibinfo{year}{1953}),
  \bibinfo{pages}{1095--1100}.
\newblock


\bibitem[Shen et~al\mbox{.}(2009a)]%
        {Shen2009a}
\bibfield{author}{\bibinfo{person}{Jianqiang Shen}, \bibinfo{person}{Erin
  Fitzhenry}, {and} \bibinfo{person}{Thomas~G Dietterich}.}
  \bibinfo{year}{2009}\natexlab{a}.
\newblock \showarticletitle{Discovering frequent work procedures from resource
  connections}. In \bibinfo{booktitle}{\emph{Proceedings of the 14th
  international conference on Intelligent user interfaces}}.
  \bibinfo{pages}{277--286}.
\newblock


\bibitem[Shen et~al\mbox{.}(2009b)]%
        {Shen2009b}
\bibfield{author}{\bibinfo{person}{Jianqiang Shen}, \bibinfo{person}{Jed
  Irvine}, \bibinfo{person}{Xinlong Bao}, \bibinfo{person}{Michael Goodman},
  \bibinfo{person}{Stephen Kolibaba}, \bibinfo{person}{Anh Tran},
  \bibinfo{person}{Fredric Carl}, \bibinfo{person}{Brenton Kirschner},
  \bibinfo{person}{Simone Stumpf}, {and} \bibinfo{person}{Thomas~G
  Dietterich}.} \bibinfo{year}{2009}\natexlab{b}.
\newblock \showarticletitle{Detecting and correcting user activity switches:
  algorithms and interfaces}. In \bibinfo{booktitle}{\emph{Proceedings of the
  14th international conference on Intelligent user interfaces}}.
  \bibinfo{pages}{117--126}.
\newblock


\bibitem[Smith and Lieberman(2010)]%
        {Smith2010}
\bibfield{author}{\bibinfo{person}{Dustin~A Smith} {and} \bibinfo{person}{Henry
  Lieberman}.} \bibinfo{year}{2010}\natexlab{}.
\newblock \showarticletitle{The why UI: using goal networks to improve user
  interfaces}. In \bibinfo{booktitle}{\emph{Proceedings of the 15th
  international conference on Intelligent user interfaces}}.
  \bibinfo{pages}{377--380}.
\newblock


\bibitem[Soh et~al\mbox{.}(2017)]%
        {soh2017deep}
\bibfield{author}{\bibinfo{person}{Harold Soh}, \bibinfo{person}{Scott Sanner},
  \bibinfo{person}{Madeleine White}, {and} \bibinfo{person}{Greg Jamieson}.}
  \bibinfo{year}{2017}\natexlab{}.
\newblock \showarticletitle{Deep sequential recommendation for personalized
  adaptive user interfaces}. In \bibinfo{booktitle}{\emph{Proceedings of the
  22nd international conference on intelligent user interfaces}}.
  \bibinfo{pages}{589--593}.
\newblock


\bibitem[Stephanidis et~al\mbox{.}(1997)]%
        {Stephanidis1997}
\bibfield{author}{\bibinfo{person}{Constantine Stephanidis},
  \bibinfo{person}{Charalampos Karagiannidis}, {and}
  \bibinfo{person}{Adamantios Koumpis}.} \bibinfo{year}{1997}\natexlab{}.
\newblock \showarticletitle{Decision making in intelligent user interfaces}. In
  \bibinfo{booktitle}{\emph{Proceedings of the 2nd international conference on
  Intelligent user interfaces}}. \bibinfo{pages}{195--202}.
\newblock


\bibitem[Su et~al\mbox{.}(2017)]%
        {Su2017}
\bibfield{author}{\bibinfo{person}{Pei-Hao Su}, \bibinfo{person}{Pawel
  Budzianowski}, \bibinfo{person}{Stefan Ultes}, \bibinfo{person}{Milica
  Gasic}, {and} \bibinfo{person}{Steve Young}.}
  \bibinfo{year}{2017}\natexlab{}.
\newblock \showarticletitle{Sample-efficient actor-critic reinforcement
  learning with supervised data for dialogue management}.
\newblock \bibinfo{journal}{\emph{arXiv preprint arXiv:1707.00130}}
  (\bibinfo{year}{2017}).
\newblock
\urldef\tempurl%
\url{https://arxiv.org/abs/1707.00130}
\showURL{%
\tempurl}


\bibitem[Sutton et~al\mbox{.}(1998)]%
        {sutton1998introduction}
\bibfield{author}{\bibinfo{person}{Richard~S Sutton}, \bibinfo{person}{Andrew~G
  Barto}, {et~al\mbox{.}}} \bibinfo{year}{1998}\natexlab{}.
\newblock \showarticletitle{Introduction to reinforcement learning}.
\newblock  (\bibinfo{year}{1998}).
\newblock


\bibitem[Tian et~al\mbox{.}(2020)]%
        {tian2020implicit}
\bibfield{author}{\bibinfo{person}{Zheng Tian}, \bibinfo{person}{Shihao Zou},
  \bibinfo{person}{Ian Davies}, \bibinfo{person}{Tim Warr},
  \bibinfo{person}{Lisheng Wu}, \bibinfo{person}{Haitham~Bou Ammar}, {and}
  \bibinfo{person}{Jun Wang}.} \bibinfo{year}{2020}\natexlab{}.
\newblock \showarticletitle{Learning to communicate implicitly by actions}. In
  \bibinfo{booktitle}{\emph{Proceedings of the AAAI Conference on Artificial
  Intelligence}}, Vol.~\bibinfo{volume}{34}. \bibinfo{pages}{7261--7268}.
\newblock


\bibitem[Todi et~al\mbox{.}(2021)]%
        {todi2021adapting}
\bibfield{author}{\bibinfo{person}{Kashyap Todi}, \bibinfo{person}{Gilles
  Bailly}, \bibinfo{person}{Luis Leiva}, {and} \bibinfo{person}{Antti
  Oulasvirta}.} \bibinfo{year}{2021}\natexlab{}.
\newblock \showarticletitle{Adapting User Interfaces with Model-based
  Reinforcement Learning}. In \bibinfo{booktitle}{\emph{Proceedings of the 2021
  CHI Conference on Human Factors in Computing Systems (CHI '21)}}.
  \bibinfo{publisher}{ACM}.
\newblock
\urldef\tempurl%
\url{https://userinterfaces.aalto.fi/adaptive/}
\showURL{%
\tempurl}


\bibitem[Wang et~al\mbox{.}(2020)]%
        {wang2020curriculum}
\bibfield{author}{\bibinfo{person}{Weixun Wang}, \bibinfo{person}{Tianpei
  Yang}, \bibinfo{person}{Yong Liu}, \bibinfo{person}{Jianye Hao},
  \bibinfo{person}{Xiaotian Hao}, \bibinfo{person}{Yujing Hu},
  \bibinfo{person}{Yingfeng Chen}, \bibinfo{person}{Changjie Fan}, {and}
  \bibinfo{person}{Yang Gao}.} \bibinfo{year}{2020}\natexlab{}.
\newblock \showarticletitle{From few to more: Large-scale dynamic multiagent
  curriculum learning}. In \bibinfo{booktitle}{\emph{Proceedings of the AAAI
  Conference on Artificial Intelligence}}, Vol.~\bibinfo{volume}{34}.
  \bibinfo{pages}{7293--7300}.
\newblock


\bibitem[Yang et~al\mbox{.}(2020)]%
        {yang2020predicting}
\bibfield{author}{\bibinfo{person}{Zhibo Yang}, \bibinfo{person}{Lihan Huang},
  \bibinfo{person}{Yupei Chen}, \bibinfo{person}{Zijun Wei},
  \bibinfo{person}{Seoyoung Ahn}, \bibinfo{person}{Gregory Zelinsky},
  \bibinfo{person}{Dimitris Samaras}, {and} \bibinfo{person}{Minh Hoai}.}
  \bibinfo{year}{2020}\natexlab{}.
\newblock \showarticletitle{Predicting goal-directed human attention using
  inverse reinforcement learning}. In \bibinfo{booktitle}{\emph{Proceedings of
  the IEEE/CVF conference on computer vision and pattern recognition}}.
  \bibinfo{pages}{193--202}.
\newblock


\bibitem[Yorke-Smith et~al\mbox{.}(2012)]%
        {Yorke2012}
\bibfield{author}{\bibinfo{person}{Neil Yorke-Smith}, \bibinfo{person}{Shahin
  Saadati}, \bibinfo{person}{Karen~L Myers}, {and} \bibinfo{person}{David~N
  Morley}.} \bibinfo{year}{2012}\natexlab{}.
\newblock \showarticletitle{The design of a proactive personal agent for task
  management}.
\newblock \bibinfo{journal}{\emph{International Journal on Artificial
  Intelligence Tools}} \bibinfo{volume}{21}, \bibinfo{number}{01}
  (\bibinfo{year}{2012}), \bibinfo{pages}{1250004}.
\newblock


\bibitem[Yu et~al\mbox{.}(2021)]%
        {yu2021surprising}
\bibfield{author}{\bibinfo{person}{Chao Yu}, \bibinfo{person}{Akash Velu},
  \bibinfo{person}{Eugene Vinitsky}, \bibinfo{person}{Yu Wang},
  \bibinfo{person}{Alexandre Bayen}, {and} \bibinfo{person}{Yi Wu}.}
  \bibinfo{year}{2021}\natexlab{}.
\newblock \showarticletitle{The surprising effectiveness of ppo in cooperative,
  multi-agent games}.
\newblock \bibinfo{journal}{\emph{arXiv preprint arXiv:2103.01955}}
  (\bibinfo{year}{2021}).
\newblock


\bibitem[Yu et~al\mbox{.}(2015)]%
        {chao2015social}
\bibfield{author}{\bibinfo{person}{Chao Yu}, \bibinfo{person}{Minjie Zhang},
  \bibinfo{person}{Fenghui Ren}, {and} \bibinfo{person}{Guozhen Tan}.}
  \bibinfo{year}{2015}\natexlab{}.
\newblock \showarticletitle{Emotional Multiagent Reinforcement Learning in
  Spatial Social Dilemmas}.
\newblock \bibinfo{journal}{\emph{IEEE Transactions on Neural Networks and
  Learning Systems}} \bibinfo{volume}{26}, \bibinfo{number}{12}
  (\bibinfo{year}{2015}), \bibinfo{pages}{3083--3096}.
\newblock
\urldef\tempurl%
\url{https://doi.org/10.1109/TNNLS.2015.2403394}
\showDOI{\tempurl}


\bibitem[Zhang et~al\mbox{.}(2021)]%
        {zhang2021multi}
\bibfield{author}{\bibinfo{person}{Kaiqing Zhang}, \bibinfo{person}{Zhuoran
  Yang}, {and} \bibinfo{person}{Tamer Ba{\c{s}}ar}.}
  \bibinfo{year}{2021}\natexlab{}.
\newblock \showarticletitle{Multi-agent reinforcement learning: A selective
  overview of theories and algorithms}.
\newblock \bibinfo{journal}{\emph{Handbook of Reinforcement Learning and
  Control}} (\bibinfo{year}{2021}), \bibinfo{pages}{321--384}.
\newblock


\end{thebibliography}

\clearpage
\appendix
\section{Curriculum Learning}
\label{app:curriculum}

We use curriculum learning for all settings. Specifically, we adjust the difficulty level every time a criteria has been met by increasing the mean number of initial attribute differences. More initial attribute differences result in longer action sequences and are therefore more complex to learn. We increase the mean by 0.01 every time the successful completion rate is above 90\% and the last level up was at least 10 epochs away.

\del{We randomly sample the number of attribute differences from a normal distribution and round it to the nearest integer. The mean of this distribution is the current task level divided by 100, and the standard deviation is $1$. we clip sampled the number of attribute differences to the range $[1, n_a]$, where $n_a$ is the number of attributes in the setting (in the case of photo editing $5$).}
\add{We randomly sample the number of attribute differences from a normal distribution with standard deviation $1$, normalize the sampled number into the range $[1, n_a]$ and round it to the nearest integer, where $n_a$ is the number of attributes of a setting (in the case of game character $n_a=5$). }

\section{Learned Lower Level}
\label{app:learnedUser}
The low-level motor control policy controls the end-effector movement. In particular, given a target slot and a speed-accuracy trade-off weight, the policy selects the parameters of an endpoint distribution. Given the current position and the endpoint parameters (mean and standard deviation), we compute the predicted movement time using the WHo Model \cite{guiard2015mathematical}. The low-level policy needs to learn i) the coordinates and dimensions of menu slots, ii) an optimal speed-accuracy trade-off given a target slot, and its current position. 

To prevent the low-level motor control policy from correcting wrong high-level decisions and to increase general performance, we limit the state space $\StatePerPolicy_M$ to strictly necessary elements with respect to the motor control task \cite{christen2021hide}: 

\begin{equation}
    \StatePerPolicy_M = \left (\pos, \target \right ),
\end{equation}
with the current position $\pos \in I^2$, the target slot $\target \in \mathbb{Z}_2^{\nslots}$. 

The action space $\ActionPerPolicy_M$ is defined as follows:

\begin{equation}
    \ActionPerPolicy_M = \left(\mu_{\pos}, \sigma_{\pos} \right).
\end{equation}
It consists of $\mu_{\pos} \in I^2$ and $\sigma_{\pos} \in I$, i.e., the mean and standard deviation which describes the endpoint distribution in the unit interval. We scale the standard deviation linearly between a min and max value where the minimum value is the size of normalized pixel width and the max value is empirically chosen to be 15\% of the screen width. Once an action is taken, we sample a new end-effector position from a normal distribution: $\pos \sim \mathcal{N}\left(\mu_{\pos}, \sigma_{\pos}\right)$.

Given the predicted actions, we compute the expected movement time via the WHo model \cite{guiard2015mathematical}, similar to our non-learned low-level motor control policy in the main paper. 

The reward for the low-level motor control policy is based on the \emph{motoric} speed-accuracy trade-off. Specifically, we penalize: i) missing the target supplied by the high-level $(\miss)$, and ii) the movement time ($\mt$). Furthermore,  we add a penalty iii)  which amounts to the squared Euclidean distance between the center of the target $\target$ and $\mu_\pos$. This incentivizes the policy to hit the desired target in the correct location. Since the penalty only considers the desired point $\mu_\pos$, it will not impact the speed-accuracy trade-off (which is a function of $\sigma_\pos$). The total reward is defined as follows:

\begin{equation}
    \RewardPerPolicy_M = \underbrace{\satweight (\miss)}_{i)} - \underbrace{(1-\satweight) \mt}_{ii)} - \underbrace{\beta ||\mu_{\pos} - \mu_\target||_2^2}_{iii)},
\end{equation}
where $\miss$ equals $0$ when the target button is hit and $-1$ on a miss. A hit occurs when the newly sampled user position $\pos$ is within the target $\target$, while a miss happens if the user position is outside of the target. $\satweight$ is a speed-accuracy trade-off weight and $\beta$ is a small scalar weight to help with learning.

\section{2D Hierarchical Menu}
\label{app:hierarchicalmenu}

\begin{figure}
    \centering
    \includegraphics[width=\columnwidth]{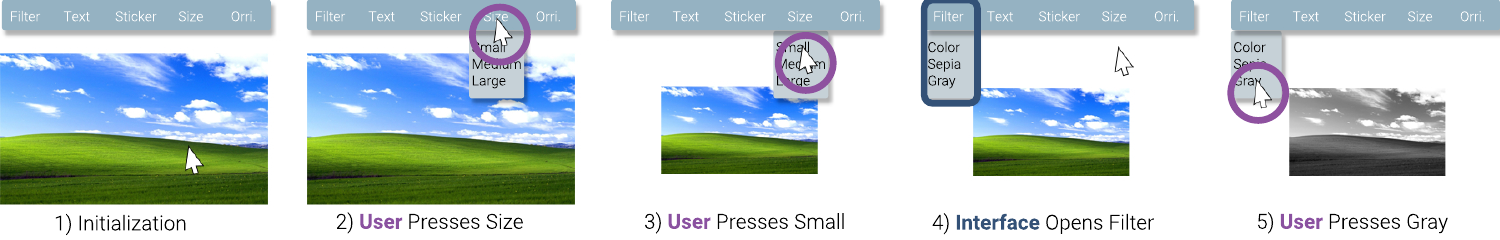}
    \caption{We introduce a photo editing task where (1)~a user matches a photo to a target by operating a hierarchical menu. (2)~The user selects the submenu `size`. (3)~The user then selects the attribute `small`, which alters the image. (4)~After the user has changed an attribute, the interface observes the new state of the photo and finds the most likely submenu for the next user action. (5)~The user clicks on an item in the submenu to complete the task.}
    \Description{
        A series of five figures that describe each step of a photo editing task is presented. A toolbar, photo, and cursor are presented in each figure, which has five components: 1) Filter, 2) Text, 3) Sticker, 4) Size, and 5) Orri., which is the abbreviation of orientation. The first figure describes initialization, where the cursor is on the photo. The second figure shows how the user opens the size menu, so the cursor is on the size button on the toolbar. The third figure shows that the user selects a small-size option by moving the cursor over the small choice. The fourth figure shows that the filter menu is opened by the interface agent. Therefore, the cursor is on the photo, not on the toolbar. The fifth and last figure describes that the user selected the grey filter from the previously opened filter menu by moving the cursor. 
    }
    \label{fig:hierarchical_good}
\end{figure}

In this task, a user edits a photo by changing its attributes. A photo has five distinct attributes with three states per attribute: i) filter (color, sepia, gray), ii) text (none, Lorem, Ipsum), iii) sticker (none, unicorn, cactus), iv) size (small, medium large), and v) orientation (original, flipped horizontal, and vertical). The photo's attribute states are limited to one per attribute, i.e., the photo cannot be in grayscale and color simultaneously. This leads to a total of 15 attribute states and 243 photo configurations. 

The graphical interface is a hierarchical menu, where each attribute is a top-level menu entry, and each attribute state is in the corresponding submenu. By clicking a top-level menu, the submenu expands and thus becomes visible and selectable. Only one menu can be expanded at any given time. 

The photo attribute states correspond to the current input state  $\tools$ and the target state $\gattr$, where $\gattr$ is only known to the \useragent. The \interfaceagent selects an attribute menu to open. Its goal is to reduce the number of clicks necessary to change an attribute, e.g., from two user interactions (filter->color) to one (color). 
For the \emph{user agent}, the higher level selects a target slot, and the lower level moves to the corresponding location.

\end{document}